\documentclass{article}\usepackage[]{graphicx}\usepackage[]{color}
\makeatletter
\def\maxwidth{ %
  \ifdim\Gin@nat@width>\linewidth
    \linewidth
  \else
    \Gin@nat@width
  \fi
}
\makeatother

\definecolor{fgcolor}{rgb}{0.345, 0.345, 0.345}
\newcommand{\hlnum}[1]{\textcolor[rgb]{0.686,0.059,0.569}{#1}}%
\newcommand{\hlstr}[1]{\textcolor[rgb]{0.192,0.494,0.8}{#1}}%
\newcommand{\hlopt}[1]{\textcolor[rgb]{0,0,0}{#1}}%
\newcommand{\hlstd}[1]{\textcolor[rgb]{0.345,0.345,0.345}{#1}}%
\newcommand{\hlkwa}[1]{\textcolor[rgb]{0.161,0.373,0.58}{\textbf{#1}}}%
\newcommand{\hlkwb}[1]{\textcolor[rgb]{0.69,0.353,0.396}{#1}}%
\newcommand{\hlkwc}[1]{\textcolor[rgb]{0.333,0.667,0.333}{#1}}%
\newcommand{\hlkwd}[1]{\textcolor[rgb]{0.737,0.353,0.396}{\textbf{#1}}}%

\usepackage{framed}
\makeatletter
\newenvironment{kframe}{%
 \def\at@end@of@kframe{}%
 \ifinner\ifhmode%
  \def\at@end@of@kframe{\end{minipage}}%
  \begin{minipage}{\columnwidth}%
 \fi\fi%
 \def\FrameCommand##1{\hskip\@totalleftmargin \hskip-\fboxsep
 \colorbox{shadecolor}{##1}\hskip-\fboxsep
     \hskip-\linewidth \hskip-\@totalleftmargin \hskip\columnwidth}%
 \MakeFramed {\advance\hsize-\width
   \@totalleftmargin\z@ \linewidth\hsize
   \@setminipage}}%
 {\par\unskip\endMakeFramed%
 \at@end@of@kframe}
\makeatother

\definecolor{shadecolor}{rgb}{.97, .97, .97}
\definecolor{messagecolor}{rgb}{0, 0, 0}
\definecolor{warningcolor}{rgb}{1, 0, 1}
\definecolor{errorcolor}{rgb}{1, 0, 0}
\newenvironment{knitrout}{}{} % an empty environment to be redefined in TeX

\usepackage{alltt}          % twocolumn
\usepackage{authblk}
\usepackage{fullpage}
\usepackage{amsthm}
\usepackage{graphicx}
\usepackage{url}
\usepackage{xcolor}
\usepackage{natbib}
\usepackage{amsmath}
\usepackage{amssymb}
\usepackage{arydshln}
\usepackage{hyperref}
\newcommand{\mb}{\mathbf}

\newcommand{\code}[1]{\texttt{#1}}
\newcommand{\proglang}[1]{\textsf{#1}}
\newcommand{\pkg}[1]{{\fontseries{b}\selectfont #1}}

\graphicspath{{./img/}}

\IfFileExists{upquote.sty}{\usepackage{upquote}}{}
\begin{document}

\title{\pkg{sgmcmc}: An \proglang{R} Package for Stochastic Gradient Markov Chain Monte Carlo}

\author{Jack Baker$^1$\footnote{email: \texttt{j.baker1@lancaster.ac.uk}}\hskip 2em 
        Paul Fearnhead$^1$\hskip 2em 
        Emily B. Fox$^2$\hskip 2em 
        Christopher Nemeth$^1$}
\affil{$^1$ STOR-i Centre for Doctoral Training, Department of Mathematics and Statistics, Lancaster University, Lancaster, UK \\ $^2$ Department of Statistics, University of Washington, Seattle, WA}

\date{}

\maketitle

\begin{abstract}
    This paper introduces the \proglang{R} package \pkg{sgmcmc}; which can be used for Bayesian inference on problems with large datasets using stochastic gradient Markov chain Monte Carlo (SGMCMC). Traditional Markov chain Monte Carlo (MCMC) methods, such as Metropolis-Hastings, are known to run prohibitively slowly as the dataset size increases. SGMCMC solves this issue by only using a subset of data at each iteration. SGMCMC requires calculating gradients of the log likelihood and log priors, which can be time consuming and error prone to perform by hand. The \pkg{sgmcmc} package calculates these gradients itself using automatic differentiation, making the implementation of these methods much easier. To do this, the package uses the software library \pkg{TensorFlow}, which has a variety of statistical distributions and mathematical operations as standard, meaning a wide class of models can be built using this framework. SGMCMC has become widely adopted in the machine learning literature, but less so in the statistics community. We believe this may be partly due to lack of software; this package aims to bridge this gap.
\end{abstract}


\section{Introduction}

This article introduces \pkg{sgmcmc}, an \proglang{R} package \citep{R2008} for scalable Bayesian inference on a wide variety of models using stochastic gradient Markov chain Monte Carlo (SGMCMC). A disadvantage of most traditional Markov chain Monte Carlo (MCMC) methods are that they require calculations over the full dataset at each iteration; meaning the methods are prohibitively slow for large datasets. SGMCMC methods provide a solution to this issue. The methods use only a subset of the full dataset, known as a minibatch, at each MCMC iteration. While the methods no longer target the true posterior, they instead produce accurate approximations to the posterior at a reduced computational cost.

The \pkg{sgmcmc} package implements a number of popular SGMCMC samplers including stochastic gradient Langevin dynamics (SGLD) \citep{Welling2011}, stochastic gradient Hamiltonian Monte Carlo (SGHMC) \citep{Chen2014} and stochastic gradient Nos\'e-Hoover thermostats (SGNHT) \citep{Ding2014}. Recent work has shown how control variates can be used to reduce the computational cost of SGMCMC algorithms \citep{{Baker2017,Nagapetyan2017}}. For each of the samplers implemented in the package, there is also a corresponding control variate sampler providing improved sampling efficiency.

Performing statistical inference on a model using SGMCMC requires calculating the gradient of the log likelihood and log priors. Calculating gradients by hand is often time consuming and error prone. One of the major advantages of \pkg{sgmcmc} is that gradients are calculated within the package using automatic differentiation \citep{Griewank2008}. This means that users need only specify the log likelihood function and log prior for their model. The package calculates the gradients using \pkg{TensorFlow} \citep{Tensorflow2015}, which has recently been made available for \proglang{R} \citep{Tensor4R}. \pkg{TensorFlow} is an efficient library for numerical computation which can take advantage of a wide variety of architectures, as such, \pkg{sgmcmc} keeps much of this efficiency. Both \pkg{sgmcmc} and \pkg{TensorFlow} are available on CRAN, so \pkg{sgmcmc} can be installed by using the standard \code{install.packages} function. Though after the \pkg{TensorFlow} package has been installed, the extra \code{install\_tensorflow()} function needs to be run, which installs the required \proglang{Python} implementation of \pkg{TensorFlow}.\footnote{More information on installing \pkg{TensorFlow} for \proglang{R} can be found at \url{https://tensorflow.rstudio.com/}.} The \pkg{sgmcmc} package also has a website with vignettes, tutorials and an API reference.\footnote{\pkg{sgmcmc} website at \url{https://stor-i.github.io/sgmcmc}}

SGMCMC methods have become popular in the machine learning literature but less so in the statistics community. We partly attribute this to the lack of available software. To the best of our knowledge, there are currently no \proglang{R} packages available for SGMCMC, probably the most popular programming language within the statistics community. The only package we are aware of which implements scalable MCMC is the \proglang{Python} package \pkg{edward} \citep{Tran2016}. This package implements both SGLD and SGHMC, but does not implement SGNHT or any of the control variate methods.

Section \ref{sec:mcmc} introduces MCMC and discusses the software currently available for implementing MCMC algorithms, we discuss the scenarios where \pkg{sgmcmc} is designed to be used. In Section \ref{sec:sgmcmc} we review the methodology behind the SGMCMC methods implemented in \pkg{sgmcmc}. Section \ref{sec:tf} provides a brief introduction to \pkg{TensorFlow}. Section \ref{sec:implementation} overviews the structure of the package, as well as details of how the algorithms are implemented. Section \ref{sec:simulations} presents a variety of example simulations. Finally, Section \ref{sec:discussion} provides a discussion on benefits and drawbacks of the implementation, as well as how we plan to extend the package in the future.

\section{Introduction to MCMC and available software}
\label{sec:mcmc}

Suppose we have a dataset of size $N$, with data $\mb x = \{x_i\}_{i=1}^N$, where $x_i \in \mathcal{X}$ for some space $\mathcal X$. We denote the probability density of $x_i$ as $p(x_i | \theta)$, where $\theta \in \Theta \subseteq \mathbb{R}^p$ are model parameters to be inferred. We assign a prior density $p(\theta)$ to the parameters and the resulting posterior is then $p( \theta | \mb x ) = p( \theta ) \prod_{i=1}^N p( x_i | \theta )/Z$.

Often the posterior can only be calculated up to a constant of proportionality $Z$. In practice $Z$ is rarely analytically tractable; so MCMC provides a way to construct a Markov chain using only the unnormalized posterior density $h(\theta) := p( \theta ) \prod_{i=1}^N p( x_i | \theta )$. The Markov chain is designed so that its stationary distribution is the posterior $p( \theta | \mb x)$. The result (once the chain has converged) is a sample $\{\theta_t\}_{t=1}^\top$ from the posterior, though this sample is not independent. A downside of these traditional MCMC algorithms is that they require the evaluation of the unnormalized density $h(\theta)$ at every iteration. This results in an $O(N)$ cost per iteration. Thus MCMC becomes prohibitively slow on large datasets.

The Metropolis-Hastings algorithm is a type of MCMC algorithm. New proposed samples $\theta^\prime$ are drawn from a proposal distribution $q(\theta^\prime|\theta)$ and then accepted with probability,
\begin{equation}
\label{eq:mh}
\min\left\{1,\frac{p(\theta^\prime|\mb x)q(\theta|\theta^\prime)}{p(\theta|\mb x)q(\theta^\prime|\theta)}\right\}.
\end{equation}
Notice that the normalising constant $Z$ cancels in \eqref{eq:mh}, so we can interchange the posterior $p(\theta | \mb x)$ with $h(\theta)$. The efficiency of the Metropolis-Hastings algorithm is dependent on the choice of proposal distribution, $q$. 

There are a number of proposals for the Metropolis-Hastings algorithm which can have a very high acceptance rate. Some examples are the Metropolis-adjusted Langevin algorithm (MALA; see e.g., \cite{Roberts1998}) and Hamiltonian Monte Carlo (HMC; see \cite{Neal2010}). The reason these proposals achieve such high acceptance rates is that they approximate a continuous diffusion process whose stationary distribution is $p(\theta|\mb x)$. As an example, consider the MALA algorithm. The MALA algorithm uses a Euler discretisation of the Langevin diffusion as the proposal,
\begin{equation}
    q(\theta' | \theta) = \mathcal N\left( \theta' \, | \, \theta + \frac{\epsilon}{2} \nabla_{\theta} \log p( \theta | \mb x ), \epsilon \mathrm I \right),
    \label{eq:mala}
\end{equation}
where $\mathcal N(\theta | \mu, \Sigma)$ denotes a multivariate normal density evaluated at $\theta$ with mean $\mu$ and variance $\Sigma$; $\mathrm I$ is simply the identity matrix; $\epsilon$ is a tuning parameter referred to as the stepsize. Discretising the diffusion introduces an approximation error, which is corrected by the Metropolis-Hastings acceptance step \eqref{eq:mh}. This means that as $\epsilon \rightarrow 0$, we tend back towards the exact, continuous diffusion and the acceptance rate is 1. However this would result in a Markov chain that never moves. Thus picking $\epsilon$ is a balance between a high acceptance rate and good mixing.

\begin{figure}[t]
    \centering
    \includegraphics[width=200px]{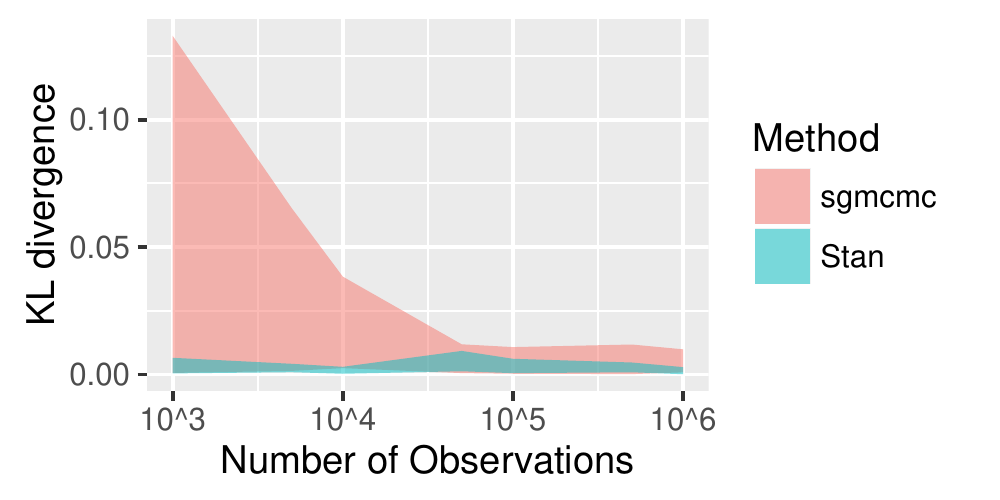}
    \includegraphics[width=200px]{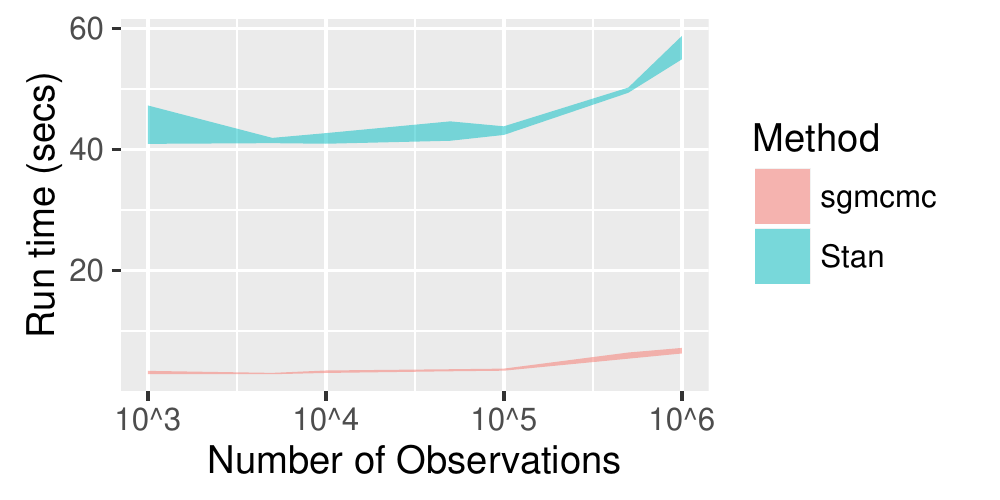}
    \caption{KL divergence (left) and run time (right) of the standard \pkg{Stan} algorithm and the \code{sgldcv} algorithm of the \pkg{sgmcmc} package when each are used to sample from data following a standard normal distribution as the number of observations are increased.}
    \label{fig:stan-sgmcmc}
\end{figure}

    There are a number of general purpose samplers for MCMC that fulfil different purposes to \pkg{sgmcmc}. The most popular samplers are \pkg{Stan}, \pkg{BUGS} and \pkg{JAGS} \citep{{Carpenter2016, Plummer2003, Lunn2000}}. The samplers \pkg{BUGS} and \pkg{JAGS} implement automated Gibbs sampling. These samplers work with both continuous and discrete parameter spaces and can be highly efficient. However because the samplers rely on Gibbs sampling, conjugate priors need to be used; also the samplers are not efficient when there is high correlation between the parameters \citep{Carpenter2016}. The package \pkg{Stan} implements state of the art HMC techniques, which means non-conjugate priors can be used, and that the sampler is more robust when there is correlation between parameters. However the package cannot perform inference on discrete parameters, and requires that these are integrated out of the model.

    The \pkg{sgmcmc} package aims to fill a gap when the dataset is large enough that other general purpose MCMC samplers such as \pkg{Stan}, \pkg{BUGS} and \pkg{JAGS} cannot be run or run prohibitively slowly. In the packages \pkg{Stan}, \pkg{BUGS} and \pkg{JAGS}, properly specified models will define a Markov chain whose stationary distribution is the posterior distribution. However a major problem with these methods are that as the number of observations gets large the algorithms run slowly. This has become a problem as dataset sizes have been increasing. The reason these methods are slow when running on large datasets are because they require a calculation over the full dataset at each iteration. The methods implemented in \pkg{sgmcmc} aim to account for this issue by only using a subset of the dataset at each iteration. The main downside being that the stationary distribution is no longer the true posterior, just a close approximation. However, as the dataset size increases, the main tuning constant in \pkg{sgmcmc}, known as the stepsize, can be set smaller and the approximation to the posterior improves. Compared to \pkg{Stan}, \pkg{BUGS} and \pkg{JAGS}; \pkg{sgmcmc} offers significant computational advantages for Bayesian modelling with large datasets, but like \pkg{Stan}, a downside of \pkg{sgmcmc} is it requires that discrete parameters are integrated out. The package also requires more tuning than other general purpose samplers since satisfactory results for tuning these methods are still under development.

    Figure \ref{fig:stan-sgmcmc} demonstrates in which scenarios practitioners may find \pkg{sgmcmc} useful. The standard \pkg{Stan} sampler and the \code{sgldcv} algorithm of the \pkg{sgmcmc} package are used to sample from the posterior of data drawn from a standard normal $N(0,1)$ with a $N(0, 10)$ prior. The Kullback-Leibler (KL) divergence between the MCMC sample and the true posterior is calculated, and the plots show how this and the run time changes as the number of observations are increased. Since both \pkg{Stan} and \pkg{TensorFlow} models need to be compiled, we recompile the models each time they are run to keep the comparison fair; but it is worth mentioning that the \pkg{Stan} run time is much quicker for the small observation models if precompiled. We can see the run time of \pkg{Stan} increasing rapidly as the number of observations is increased, while the run time of the \pkg{sgmcmc} algorithm increases more slowly. This is a very simple model, used so that the KL divergence can be calculated exactly. As the model complexity increases the run time of \pkg{Stan} can quickly become unmanageable for large datasets. On the other hand, we can see that the KL divergence of the \pkg{sgmcmc} algorithm for this example is poor compared with \pkg{Stan} for a small number of observations. However, as the dataset size increases, the KL divergence for the \pkg{sgmcmc} algorithm becomes much more reasonable compared with \pkg{Stan}. Thus \pkg{sgmcmc} is best used when the dataset size is slowing down the run time of the other general purpose algorithms, and practitioners can safely trade-off a small amount of accuracy in order to gain significant speed-ups.

\section{Stochastic gradient MCMC}
\label{sec:sgmcmc}

Many popular MCMC proposal distributions, including HMC and MALA, described in \eqref{eq:mala}, require the calculation of the gradient of the log posterior at each iteration, which is an $O(N)$ calculation. Stochastic gradient MCMC methods get around this by replacing the true gradient with the following unbiased estimate
\begin{equation}
    \widehat{\nabla_{\theta} \log p( \theta_t | \mb x ) } := \nabla_{\theta} \log p(\theta_t) + \frac{N}{n} \sum_{i \in S_t} \nabla_{\theta} \log p( x_i | \theta_t ),
    \label{eq:std-grad}
\end{equation}
calculated on some subset of the all observations $S_t \subset \{1,\dots,N\}$, known as a minibatch, with $|S_t| = n$.

Calculating the Metropolis-Hastings acceptance step \eqref{eq:mh} is another $O(N)$ calculation. To get around this, SGMCMC methods set the tuning constants such that the acceptance rate will be high, and remove the acceptance step from the algorithm altogether. By ignoring the acceptance step, and estimating the log posterior gradient, the per iteration cost of SGMCMC algorithms is $O(n)$, where $n$ is the minibatch size. However, the algorithm no longer targets the true posterior but an approximation. There has been a body of theory exploring how these methods perform in different settings. Similar to MALA, the algorithms rely on a stepsize parameter $\epsilon$. Some of the algorithms have been shown to weakly converge as $\epsilon \rightarrow 0$.

\subsection{Stochastic gradient Langevin dynamics}
\label{sec:sgld-review}

SGLD, first introduced by \cite{Welling2011}, is an SGMCMC approximation to the MALA algorithm. By substituting \eqref{eq:std-grad} into the MALA proposal equation \eqref{eq:mala}, we arrive at the following update for $\theta$
\begin{equation}
    \theta_{t+1} = \theta_t + \frac{\epsilon_t}{2} \widehat{ \nabla_{\theta} \log p( \theta_t | \mb x ) } + \zeta_t,
\end{equation}
where $\zeta_t \sim \mathcal{N}(0,\epsilon_t)$.

\cite{Welling2011} noticed that as $\epsilon_t \rightarrow 0$ this update will sample from the true posterior. Although the algorithm mixes slower as $\epsilon$ gets closer to 0, so setting the stepsize is a trade-off between convergence speed and accuracy. This motivated \cite{Welling2011} to suggest setting $\epsilon_t$ to decrease to 0 as $t$ increases. There is a body of theory that investigates the SGLD approximation to the true posterior \citep[see e.g.,][]{Teh2014, Sato2014, Vollmer2015}. In particular, SGLD is found to converge weakly to the true posterior distribution asymptotically as $\epsilon_t \rightarrow 0$. The mean squared error of the algorithm is found to decrease at best with rate $O(T^{-1 / 3})$. In practice, the algorithm tends to mix poorly when $\epsilon$ is set to decrease to 0 \citep{Vollmer2015}, so in our implementation we use a fixed stepsize which has been shown to mix better empirically. Theoretical analysis for this case is provided in \cite{Vollmer2015}. The tuning constant $\epsilon$, referred to as the stepsize is a required argument in the package. It affects the performance of the algorithm considerably.

\subsection{Stochastic gradient Hamiltonian Monte Carlo}
\label{sec:sghmc}

The stochastic gradient Hamiltonian Monte Carlo algorithm (SGHMC) \citep{Chen2014} is similar to SGLD, but instead approximates the HMC algorithm \citep{Neal2010}. To ensure SGHMC is $O(n)$, the same unbiased estimate to the log posterior gradient is used \eqref{eq:std-grad}. SGHMC augments the parameter space with momentum variables $\nu$ and samples approximately from a joint distribution $p( \theta, \nu | \mb x )$, whose marginal distribution for $\theta$ is the posterior of interest. The algorithm performs the following updates at each iteration
\begin{align*}
        \theta_{t+1} &= \theta_t + \nu_t, \\
        \nu_{t+1} &= (1 - \alpha) \nu_t + \epsilon \widehat{\nabla_{\theta} \log p( \theta_{t+1} | \mb x )} + \zeta_t,
\end{align*}
where $\zeta_t \sim \mathcal{N}( 0, 2[ \alpha - \hat \beta_t ] \epsilon )$; $\epsilon$ and $\alpha$ are tuning constants and $\hat \beta_t$ is proportional to an estimate of the Fisher information matrix. In our current implementation, we simply set $\hat \beta_t := 0$, as in the experiments of the original paper by \cite{Chen2014}. In future implementations, we aim to estimate $\hat \beta_t$ using a Fisher scoring estimate similar to \cite{Ahn2012}. Often the dynamics are simulated repeatedly $L$ times before the state is stored, at which point $\nu$ is resampled. The parameter $L$ can be included in our implementation. The tuning constant $\epsilon$ is the stepsize and is a required argument in our implementation, as for SGLD its value affects performance considerably. The constant $\alpha$ tends to be fixed at a small value in practice. As a result, in our implementation it is an optional argument with default value 0.01.

\subsection{Stochastic gradient Nos\'e-Hoover thermostat}

\cite{Ding2014} suggest that the quantity $\hat \beta_t$ in SGHMC is difficult to estimate in practice. They appeal to analogues between these proposals and statistical physics in order to suggest a set of updates which do not need this estimation to be made. Once again \cite{Ding2014} augment the space with momentum parameters $\nu$. They replace the tuning constant $\alpha$ with a dynamic parameter $\alpha_t$ known as the thermostat parameter. The algorithm performs the following updates at each iteration
\begin{align}
    \theta_{t+1} &= \theta_t + \nu_t, \\
    \nu_{t+1} &= (1-\alpha_t) \nu_t + \epsilon \widehat{\nabla_{\theta} \log p( \theta_{t+1} | \mb x )} + \zeta_t, \\
    \alpha_{t+1} &= \alpha_{t} + \left[ \frac 1 p (\nu_{t+1})^\top (\nu_{t+1}) - \epsilon \right].
    \label{eq:temp}
\end{align}
Here $\zeta_t \sim \mathcal{N}(0,2a\epsilon)$; $\epsilon$ and $a$ are tuning parameters to be chosen and $p$ is the dimension of $\theta$. The update for $\alpha$ in \eqref{eq:temp} requires a vector dot product, it is not obvious how to extend this when $\theta$ is higher order than a vector, such as a matrix or tensor. In our implementation, when $\theta$ is a matrix or tensor we use the standard inner product in those spaces \citep{Abraham1988}. The tuning constant $\epsilon$ is the stepsize and is a required argument in our implementation, as again its value affects performance considerably. The constant $a$, similarly to $\alpha$ in SGHMC, tends to be fixed at a small value in practice \citep{Ding2014}. As a result, in our implementation it is an optional argument with default value 0.01.

\subsection{Stochastic gradient MCMC with control variates}
\label{sec:sgmcmccv}

A key feature of SGMCMC methods is replacing the log posterior gradient calculation with an unbiased estimate. The unbiased gradient estimate, which can be viewed as a noisy version of the true gradient, can have high variance when estimated using a small minibatch of the data. Increasing the minibatch size will reduce the variance of the gradient estimate, but increase the per iteration computational cost of the SGMCMC algorithm. Recently control variates \citep{Ripley2009} have been used to reduce the variance in the gradient estimate of SGMCMC \citep{{Dubey2016,Nagapetyan2017,Baker2017}}. Using these improved gradient estimates have been shown to lead to improvements in the mean squared error (MSE) of the algorithm \citep{Dubey2016}, as well as its computational cost \citep{{Nagapetyan2017,Baker2017}}.

We implement the formulation of \cite{Baker2017}, who replace the gradient estimate $\widehat{ \nabla_{\theta} \log p( \theta | \mb x )}$ with
\begin{equation}
    \widetilde{\nabla_{\theta} \log p(\theta | \mb x )} := \nabla_{\theta} \log p( \hat \theta | \mb x ) + \widehat{ \nabla_{\theta} \log p( \theta | \mb x ) } - \widehat{ \nabla_{\theta} \log p( \hat \theta | \mb x ) },
    \label{eq:cv-grad}
\end{equation}
where $\hat \theta$ is an estimate of the posterior mode. This method requires the burn-in phase of MCMC to be replaced by an optimisation step which finds $\hat \theta := \mathrm{argmax}_\theta \log p(\theta | \mb x )$. There is then an $O(N)$ preprocessing step to calculate $\nabla_{\theta} \log p( \hat \theta | \mb x )$, after which the chain can be started from $\hat \theta$ resulting in a negligible mixing time. \cite{Baker2017} and \cite{Nagapetyan2017} have shown that there are considerable improvements to the computational cost of SGLD when \eqref{eq:cv-grad} is used in place of \eqref{eq:std-grad}. In particular they showed that standard SGLD requires setting the minibatch size $n$ to be $O(N)$ for arbitrarily good performance; while using control variates requires an $O(N)$ preprocessing step, but after that a batch size of $O(1)$ can be used to reach the desired performance. \cite{Baker2017} also showed empirically that this particular formulation can lead to a reduction in the burn-in time compared with standard SGLD and the formulation of \citep{Dubey2016}, which tended to get stuck in local stationary points. This is because in complex scenarios the optimisation step is often faster than the burn-in time of SGMCMC. The package \pkg{sgmcmc} includes control variate versions of all the SGMCMC methods implemented: SGLD, SGHMC and SGNHT.

\section[Brief TensorFlow introduction]{Brief \pkg{TensorFlow} introduction}
\label{sec:tf}

\pkg{TensorFlow} \citep{Tensorflow2015} is a software library released by Google. The tool was initially designed to easily build deep learning models; but the efficient design and excellent implementation of automatic differentiation \citep{Griewank2008} has made it useful more generally. This package is built on \pkg{TensorFlow}, and while we have tried to make the package as easy as possible to use, some knowledge of \pkg{TensorFlow} is unavoidable; especially when declaring the log likelihood and log prior functions, or for high dimensional chains which will not fit into memory. With this in mind, we provide a brief introduction to \pkg{TensorFlow} in this section. This should provide enough knowledge for the rest of the article. A more detailed introduction to \pkg{TensorFlow} for \proglang{R} can be found at \cite{Tensor4R}.

Any procedure written in \pkg{TensorFlow} follows three main steps. The first step is to declare all the variables, constants and equations required to run the algorithm. In the background, these declarations enable \pkg{TensorFlow} to build a graph of the possible operations, and how they are related to other variables, constants and operations. Once everything has been declared, the \pkg{TensorFlow} session is begun and all the variables and operations are initialised. Then the previously declared operations can be run as required; typically these will be sequential and will be run in a for loop.

\subsection[Declaring TensorFlow tensors]{Declaring \pkg{TensorFlow} tensors}

Everything in \pkg{TensorFlow}, including operations, are represented as a tensor; which is basically a multi-dimensional array. There are a number of ways of creating tensors:
\begin{itemize}
    \item \code{tf\$constant(value)} -- create a constant tensor with the same shape and values as \code{value}. The input \code{value} is generally an \proglang{R} array, vector or scalar. The most common use for this in the context of the package is for declaring constant parameters when declaring log likelihood and log prior functions.
    \item \code{tf\$Variable(value)} -- create a tensor with the same shape and values as \code{value}. Unlike \code{tf\$constant}, this type of tensor can be changed by a declared operation. The input \code{value} is generally an \proglang{R} array, vector or scalar.
    \item \code{tf\$placeholder(datatype, shape)} -- create an empty tensor of type \code{datatype} and dimensions \code{shape} which can be fed all or part of a dataset, this is useful when declaring operations which rely on data which can change. When you have storage constraints (see Section \ref{sec:nnusage}) you can use a placeholder to declare test functions that rely on a test dataset. The \code{datatype} should be a \pkg{TensorFlow} data type, such as \code{tf\$float32}; the \code{shape} should be an \proglang{R} vector or scalar, such as \code{c(100,2)}.
    \item operation -- an operation declares an operation on previously declared tensors. These use \pkg{TensorFlow} defined functions, such as those in its math library. This is essentially what you are declaring when coding the \code{logLik} and \code{logPrior} functions. The \code{params} input consists of a list of \code{tf\$Variables}, representing the model parameters to be inferred. The \code{dataset} input consists of a list of \code{tf\$placeholder} tensors, representing the minibatch of data fed at each iteration. Your job is to declare functions that return an operation on these tensors that define the log likelihood and log prior.
\end{itemize}

\subsection[TensorFlow operations]{\pkg{TensorFlow} operations}

\pkg{TensorFlow} operations take other tensors as input and manipulate them to reach the desired output. Once the \pkg{TensorFlow} session has begun, these operations can be run as needed, and will use the current values for its input tensors. For example, we could declare a normal density tensor which manipulates a \code{tf\$Variable} tensor representing the parameters and a \code{tf\$placeholder} tensor representing the current data point. The tensor could then use the \code{TensorFlow} \code{tf\$contrib\$distributions\$MultivariateNormalDiag} object to return a tensor object of the current value for a normal density given the current parameter value and the data point that's fed to the placeholder. We can break this example down into three steps. First we declare the tensors that we require:
\begin{knitrout}
\definecolor{shadecolor}{rgb}{0.969, 0.969, 0.969}\color{fgcolor}\begin{kframe}
\begin{alltt}
\hlkwd{library}\hlstd{(}\hlstr{"tensorflow"}\hlstd{)}
\hlstd{loc} \hlkwb{=} \hlstd{tf}\hlopt{$}\hlkwd{Variable}\hlstd{(}\hlkwd{rep}\hlstd{(}\hlnum{0}\hlstd{,} \hlnum{2}\hlstd{))}
\hlstd{dataPoint} \hlkwb{=} \hlstd{tf}\hlopt{$}\hlkwd{placeholder}\hlstd{(tf}\hlopt{$}\hlstd{float32,} \hlkwd{c}\hlstd{(}\hlnum{2}\hlstd{))}
\hlstd{scaleDiag} \hlkwb{=} \hlstd{tf}\hlopt{$}\hlkwd{constant}\hlstd{(}\hlkwd{c}\hlstd{(}\hlnum{1}\hlstd{,} \hlnum{1}\hlstd{))}
\hlstd{distn} \hlkwb{=} \hlstd{tf}\hlopt{$}\hlstd{contrib}\hlopt{$}\hlstd{distributions}\hlopt{$}\hlkwd{MultivariateNormalDiag}\hlstd{(loc, scaleDiag)}
\hlstd{dens} \hlkwb{=} \hlstd{distn}\hlopt{$}\hlkwd{prob}\hlstd{(dataPoint)}
\end{alltt}
\end{kframe}
\end{knitrout}

Here we have declared a \code{tf\$Variable} tensor to hold the location parameter; a \code{tf\$placeholder} tensor which will be fed the current data point; the scale parameter is fixed so we declare this as a \code{tf\$constant} tensor. Next we declare the operation which takes the inputs we just declared and returns the normal density value. The first line which is assigned to \code{distn} creates a \code{MultivariateNormalDiag} object, which is linked to the \code{loc} and \code{scaleDiag} tensors. Then \code{dens} evaluates the density of this distribution. The \code{dens} variable is now linked to the tensors \code{dataPoint} and \code{scaleDiag}, so if it is evaluated it will use the current values of those tensors to calculate the density estimate. Next we begin the \pkg{TensorFlow} session:

\begin{knitrout}
\definecolor{shadecolor}{rgb}{0.969, 0.969, 0.969}\color{fgcolor}\begin{kframe}
\begin{alltt}
\hlstd{sess} \hlkwb{=} \hlstd{tf}\hlopt{$}\hlkwd{Session}\hlstd{()}
\hlstd{sess}\hlopt{$}\hlkwd{run}\hlstd{(tf}\hlopt{$}\hlkwd{global_variables_initializer}\hlstd{())}
\end{alltt}
\end{kframe}
\end{knitrout}

The two lines we just ran starts the \pkg{TensorFlow} session and initialises all the tensors we just declared. The \pkg{TensorFlow} graph has now been built and no new tensors can now be added. This means that all operations need to be declared before they can be evaluated. Now the session is started we can run the operation \code{dens} we declared given current values for \code{dataPoint} and \code{loc} as follows:

\begin{knitrout}
\definecolor{shadecolor}{rgb}{0.969, 0.969, 0.969}\color{fgcolor}\begin{kframe}
\begin{alltt}
\hlstd{x} \hlkwb{=} \hlkwd{rnorm}\hlstd{(}\hlnum{2}\hlstd{)}
\hlstd{out} \hlkwb{=} \hlstd{sess}\hlopt{$}\hlkwd{run}\hlstd{(dens,} \hlkwc{feed_dict} \hlstd{=} \hlkwd{dict}\hlstd{(} \hlkwc{dataPoint} \hlstd{= x ) )}
\hlkwd{print}\hlstd{(}\hlkwd{paste0}\hlstd{(}\hlstr{"Density value for x is "}\hlstd{, out))}
\end{alltt}
\end{kframe}
\end{knitrout}

Since \code{dataPoint} is a \code{placeholder}, we need to feed it values each time. In the block of code above we feed \code{dataPoint} a random value simulated from a standard normal. The \code{sess\$run} expression then evaluates the current normal density value given \code{loc} and \code{dataPoint}.

As mentioned earlier, this is essentially what is happening when you are writing the \code{logLik} and \code{logPrior} functions. These functions will be fed a list of \code{tf\$Variable} objects to the \code{params} input, and a list of \code{tf\$placeholder} objects to the \code{dataset} input. The output of the function will then be declared as a \code{TensorFlow} operation. This allows the gradient to be calculated automatically, while maintaining the efficiencies of \pkg{TensorFlow}.

\pkg{TensorFlow} implements a lot of useful functions to make building these operations easier. For example a number of distributions are implemented at \code{tf\$contrib\$distributions},\footnote{For full API details see \url{https://www.tensorflow.org/api_docs/python/tf/contrib/distributions} though note this is for \proglang{Python}, so the \code{.} object notation needs to be replaced by \code{\$}, for example \code{tf.contrib.distributions.Normal} would be replaced by \code{tf\$contrib\$distributions\$Normal}.} (e.g., \code{tf\$contrib\$distributions\$Normal} and \code{tf\$contrib\$distributions\$Gamma}). \pkg{TensorFlow} also has a comprehensive math library which provides a variety of useful tensor operations.\footnote{See \url{https://www.tensorflow.org/api_guides/python/math_ops}.} For examples of writing \pkg{TensorFlow} operations see the worked examples in Section \ref{sec:implementation} or the \pkg{sgmcmc} vignettes.

\section{Package structure and implementation}
\label{sec:implementation}

The package has 6 main functions. The first three: \code{sgld}, \code{sghmc} and \code{sgnht} will implement SGLD, SGHMC and SGNHT, respectively. The other three: \code{sgldcv}, \code{sghmccv} and \code{sgnhtcv} implement the control variate versions of SGLD, SGHMC and SGNHT, respectively. All of these are built on the \pkg{TensorFlow} library for \proglang{R}, which enables gradients to be automatically calculated and efficient computations to be performed in high dimensions. The usage for these functions is very similar, with the only differences in input being tuning parameters. These main functions are outlined in Table \ref{tab:main-fns}

\begin{table}[t]
\begin{tabular}{l|l}
\hline
Function              & Algorithm \\
\hline
\code{sgld}    & Stochastic gradient Langevin dynamics \\
\code{sghmc}   & Stochastic gradient Hamiltonian Monte Carlo \\
\code{sgnht}   & Stochastic gradient Nos\'e-Hoover thermostat \\
\hdashline
\code{sgldcv}  & Stochastic gradient Langevin dynamics with control variates\\
\code{sghmccv} & Stochastic gradient Hamiltonian Monte Carlo with control variates\\
\code{sgnhtcv} & Stochastic gradient Nos\'e-Hoover thermostat with control variates\\
\hline
\end{tabular}
\caption{Outline of 6 main functions implemented in \pkg{sgmcmc}.}
\label{tab:main-fns}
\end{table}

The functions \code{sgld}, \code{sghmc} and \code{sgnht} have the same required inputs: \code{logLik}, \code{dataset}, \code{params} and \code{stepsize}. These determine respectively: the log likelihood function for the model; the data for the model; the parameters of the model; and the stepsize tuning constants for the SGMCMC algorithm. The input \code{logLik} is a function taking \code{dataset} and \code{params} as input, while the rest are defined as lists. Using lists in this way provides a lot of flexibility: allowing multiple parameters to be defined; use multiple datasets; and use different stepsizes for each parameter, which is vital if the scalings are different. It also allows users to easily reference parameters and datasets in the \code{logLik} function by simply referring to the relevant names in the list.

\begin{table}[t]
\begin{tabular}{l|l}
\hline
Function inputs              & Definition \\
\hline
\code{logLik} & Log-likelihood function taking \code{dataset} and \code{params} as inputs\\
\code{dataset} & \proglang{R} list containing data\\
\code{params} & \proglang{R} list containing model parameters\\
\code{stepsize} & \proglang{R} list of stepsizes for each parameter\\
\hdashline
\code{optStepsize} & \proglang{R} numeric stepsize for control variate optimisation step \\
\hdashline
\code{logPrior} & Function of the parameters; default $p(\theta)\propto 1$\\
\code{minibatchSize} & Size of minibatch per iteration as integer or proportion; default 0.01.\\
\code{nIters} & Number of MCMC iterations; default is 10,000.\\
\hline
\end{tabular}
\caption{Outline of the key arguments required by the functions in Table \ref{tab:main-fns}.}
\label{tab:reqd-args}
\end{table}

The functions also have a couple of optional parameters that are particularly important, \code{logPrior} and \code{minibatchSize}. These respectively define the log prior for the model; and the minibatch size, as it was defined in Section \ref{sec:sgmcmc}. By default, the \code{logPrior} is set to an uninformative uniform prior, which is fine to use for quick checks but will need to be set properly for more complex models. The \code{logPrior} is a function similar to \code{logLik}, but only takes \code{params} as input. The \code{minibatchSize} is a numeric, and can either be a proportion of the dataset size, if it is set between 0 and 1, or an integer. The default value is 0.01, meaning that $1\%$ of the full dataset is used at each iteration.

The control variate functions have the same inputs as the non-control variate functions, with one more required input. The \code{optStepsize} input is a numeric that specifies the stepsize for the initial optimisation step to find the $\hat \theta$ maximising the posterior as defined in Section \ref{sec:sgmcmccv}. For a full outline of the key inputs, see Table \ref{tab:reqd-args}.

Often large datasets and high dimensional problems go hand in hand. In these high dimensional settings storing the full MCMC chain in memory can become an issue. For this situation we provide functionality to run the chain one iteration at a time in a user defined loop. This enables the user to deal with the output at each iteration how they see fit. For example, they may wish to calculate a test function on the output to reduce the dimensionality of the chain; or they might calculate the required Monte Carlo estimates on the fly. We aim to extend this functionality to enable the user to define their own Gibbs updates alongside the SGMCMC procedure. This functionality is more involved, and requires more knowledge of \pkg{TensorFlow}, so we leave the details to the example in Section \ref{sec:nnusage}.

As mentioned earlier, \pkg{TensorFlow} already has a variety of distributions implemented. These are located in two places: \code{tf\$distributions} contains some common univariate distributions, while \code{tf\$contrib\$distributions} contains common multivariate distributions and more univariate distributions. Specifying the \code{logLik} and \code{logPrior} functions will require you to regularly specify distributions. Therefore, in the interest of cleanliness, we make the minor change of copying the distributions from \code{tf\$contrib\$distributions} to \code{tf\$distributions}. The module \code{tf\$contrib\$distributions} will still work as expected. The copying across occurs as soon as the package is loaded into the workspace. This means any distribution \code{distn} in the \pkg{TensorFlow} package can be accessed by referencing \code{tf\$distributions\$distn}.

For the rest of this section we go into more detail about the usage of the functions using a worked example. The package is used to infer the bias and coefficient parameters in a logistic regression model. Section \ref{sec:usage} demonstrates standard usage by performing inference on the model using the \code{sgld} and \code{sgldcv} functions. Section \ref{sec:nnusage} demonstrates usage in problems where the full MCMC chain cannot be fit into memory. The same logistic regression model is used throughout.

\subsection{Example usage}
\label{sec:usage}

In this example we use the functions \code{sgld} and \code{sgldcv} to infer the bias (or intercept) and coefficients of a logistic regression model. The \code{sgldcv} case is also available as a vignette. Suppose we have data $\mb x_1, \dots, \mb x_N$ of dimension $d$ taking values in $\mathbb R^d$; and response variables $y_1, \dots, y_N$ taking values in $\{0, 1\}$. Then a logistic regression model with coefficients $\beta$ and bias $\beta_0$ will have likelihood
\begin{equation}
  \label{eq:logistic}
    p( \mb X, \mb y | \beta, \beta_0 ) = \prod_{i=1}^N \left[ \frac{1}{1 + e^{-\beta_0 - \mb x_i \beta}} \right]^{y_i} \left[ 1 - \frac{1}{1 + e^{-\beta_0 - \mb x_i \beta}} \right]^{1-y_i}
\end{equation}

We will use the \code{covertype} dataset \citep{Blackard1999} which can be downloaded and loaded using the \pkg{sgmcmc} function \code{getDataset}, which downloads example datasets for the package. The \code{covertype} dataset uses mapping data to predict the type of forest cover. Our particular dataset is taken from the LIBSVM library \citep{Chang2011}, which converts the data to a binary problem, rather than multiclass. The dataset consists of a matrix of dimension $581012 \times 55$. The first column contains the labels $\mb y$, taking values in $\{0,1\}$. The remaining columns are the explanatory variables $\mb X$, which have been scaled to take values in $[0,1]$.
\begin{knitrout}
\definecolor{shadecolor}{rgb}{0.969, 0.969, 0.969}\color{fgcolor}\begin{kframe}
\begin{alltt}
\hlkwd{library}\hlstd{(}\hlstr{"sgmcmc"}\hlstd{)}
\hlstd{covertype} \hlkwb{=} \hlkwd{getDataset}\hlstd{(}\hlstr{"covertype"}\hlstd{)}
\end{alltt}
\end{kframe}
\end{knitrout}
Now we'll split the dataset into predictors and response and get the dataset into the required \proglang{R} list format.
\begin{knitrout}
\definecolor{shadecolor}{rgb}{0.969, 0.969, 0.969}\color{fgcolor}\begin{kframe}
\begin{alltt}
\hlstd{X} \hlkwb{=} \hlstd{covertype[,}\hlnum{2}\hlopt{:}\hlkwd{ncol}\hlstd{(covertype)]}
\hlstd{y} \hlkwb{=} \hlstd{covertype[,}\hlnum{1}\hlstd{]}
\hlstd{dataset} \hlkwb{=} \hlkwd{list}\hlstd{(} \hlstr{"X"} \hlstd{= X,} \hlstr{"y"} \hlstd{= y )}
\end{alltt}
\end{kframe}
\end{knitrout}
In the last line we defined the dataset as a list object which will be input to the relevant \pkg{sgmcmc} function. The list names can be arbitrary, but must be consistent with the variables declared in the \code{logLik} function (see below).

When accessing the data, it is assumed that observations are split along the first axis. In other words, \code{dataset\$X} is a 2-dimensional matrix, and we assume that observation $\mb x_i$ is accessed at \code{dataset\$X[i,]}. Similarly, suppose \code{Z} was a 3-dimensional array, we would assume that observation $i$ would be accessed at \code{Z[i,{},]}. Parameters are declared in a similar way, except the list entries are the desired parameter starting points. There are two parameters for this model, the bias $\beta_0$ and the coefficients $\beta$, which can be arbitrarily initialised to $0$.
\begin{knitrout}
\definecolor{shadecolor}{rgb}{0.969, 0.969, 0.969}\color{fgcolor}\begin{kframe}
\begin{alltt}
\hlstd{d} \hlkwb{=} \hlkwd{ncol}\hlstd{(dataset}\hlopt{$}\hlstd{X)}
\hlstd{params} \hlkwb{=} \hlkwd{list}\hlstd{(} \hlstr{"bias"} \hlstd{=} \hlnum{0}\hlstd{,} \hlstr{"beta"} \hlstd{=} \hlkwd{matrix}\hlstd{(} \hlkwd{rep}\hlstd{(} \hlnum{0}\hlstd{, d ),} \hlkwc{nrow} \hlstd{= d ) )}
\end{alltt}
\end{kframe}
\end{knitrout}

The log likelihood is specified as a function of the \code{dataset} and \code{params}, which are lists with the same names we declared earlier. The only difference is that the objects inside the lists will have automatically been converted to \pkg{TensorFlow} objects. The \code{dataset} list will contain \pkg{TensorFlow} placeholders. The \code{params} list will contain \pkg{TensorFlow} variables. The \code{logLik} function should be a function that takes these lists as input and returns the log likelihood value given the current parameters and data. This is done using \pkg{TensorFlow} operations, as this allows the gradient to be automatically calculated.

For the logistic regression model \eqref{eq:logistic}, the log likelihood is  
\[
    \log p( \mb X, \mb y | \beta, \beta_0 ) = \sum_{i=1}^N y_i\log y_{\mathrm{est}} + (1-y_i)\log(1-y_{\mathrm{est}}),
\]
where $y_{\mathrm{est}} = [1 + e^{-\beta_0 - \mb x_i \beta}]^{-1}$, which coded as a \code{logLik} function is defined as follows
\begin{knitrout}
\definecolor{shadecolor}{rgb}{0.969, 0.969, 0.969}\color{fgcolor}\begin{kframe}
\begin{alltt}
\hlstd{logLik} \hlkwb{=} \hlkwa{function}\hlstd{(}\hlkwc{params}\hlstd{,} \hlkwc{dataset}\hlstd{) \{}
  \hlstd{yEst} \hlkwb{=} \hlnum{1} \hlopt{/} \hlstd{(}\hlnum{1} \hlopt{+} \hlstd{tf}\hlopt{$}\hlkwd{exp}\hlstd{(} \hlopt{-} \hlstd{tf}\hlopt{$}\hlkwd{squeeze}\hlstd{(params}\hlopt{$}\hlstd{bias} \hlopt{+} \hlstd{tf}\hlopt{$}\hlkwd{matmul}\hlstd{(}
    \hlstd{dataset}\hlopt{$}\hlstd{X, params}\hlopt{$}\hlstd{beta))))}
  \hlstd{logLik} \hlkwb{=} \hlstd{tf}\hlopt{$}\hlkwd{reduce_sum}\hlstd{(dataset}\hlopt{$}\hlstd{y} \hlopt{*} \hlstd{tf}\hlopt{$}\hlkwd{log}\hlstd{(yEst)} \hlopt{+}
    \hlstd{(}\hlnum{1} \hlopt{-} \hlstd{dataset}\hlopt{$}\hlstd{y)} \hlopt{*} \hlstd{tf}\hlopt{$}\hlkwd{log}\hlstd{(}\hlnum{1} \hlopt{-} \hlstd{yEst))}
  \hlkwd{return}\hlstd{(logLik)}
\hlstd{\}}
\end{alltt}
\end{kframe}
\end{knitrout}
For more information about the usage of these \pkg{TensorFlow} functions, please see the \pkg{TensorFlow} documentation.\footnote{Documentation for \pkg{TensorFlow} for \proglang{R} available at \url{https://tensorflow.rstudio.com/tensorflow/}}

Next, the log prior density, where we assume each $\beta_j$, for $j = 0, \dots, d$, has an independent Laplace prior with location 0 and scale 1, so $\log p( \beta ) \propto - \sum_{j=0}^d | \beta_j |$. Similar to \code{logLik}, this is defined as a function, but with only \code{params} as input
\begin{knitrout}
\definecolor{shadecolor}{rgb}{0.969, 0.969, 0.969}\color{fgcolor}\begin{kframe}
\begin{alltt}
\hlstd{logPrior} \hlkwb{=} \hlkwa{function}\hlstd{(}\hlkwc{params}\hlstd{) \{}
  \hlstd{logPrior} \hlkwb{=} \hlopt{-} \hlstd{(tf}\hlopt{$}\hlkwd{reduce_sum}\hlstd{(tf}\hlopt{$}\hlkwd{abs}\hlstd{(params}\hlopt{$}\hlstd{beta))} \hlopt{+}
    \hlstd{tf}\hlopt{$}\hlkwd{reduce_sum}\hlstd{(tf}\hlopt{$}\hlkwd{abs}\hlstd{(params}\hlopt{$}\hlstd{bias)))}
  \hlkwd{return}\hlstd{(logPrior)}
\hlstd{\}}
\end{alltt}
\end{kframe}
\end{knitrout}

The final input that needs to be set is the \code{stepsize} for tuning the methods, this can be set as a list 
\begin{knitrout}
\definecolor{shadecolor}{rgb}{0.969, 0.969, 0.969}\color{fgcolor}\begin{kframe}
\begin{alltt}
\hlstd{stepsize} \hlkwb{=} \hlkwd{list}\hlstd{(}\hlstr{"beta"} \hlstd{=} \hlnum{2e-5}\hlstd{,} \hlstr{"bias"} \hlstd{=} \hlnum{2e-5}\hlstd{)}
\end{alltt}
\end{kframe}
\end{knitrout}

Setting the same stepsize for all parameters is done as \code{stepsize = 2e-5}. This shorthand can also be used for any of the optional tuning parameters which need to specified as lists. The stepsize parameter will generally require a bit of tuning in order to get good performance, for this we recommend using cross validation \citep[see e.g.,][Chapter 7]{Friedman2001}.

Everything is now ready to run a standard SGLD algorithm, with \code{minibatchSize} set to 500. To keep things reproducible we'll set the seed to 13.
\begin{knitrout}
\definecolor{shadecolor}{rgb}{0.969, 0.969, 0.969}\color{fgcolor}\begin{kframe}
\begin{alltt}
\hlstd{output} \hlkwb{=} \hlkwd{sgld}\hlstd{( logLik, dataset, params, stepsize,} \hlkwc{logPrior} \hlstd{= logPrior,}
  \hlkwc{minibatchSize} \hlstd{=} \hlnum{500}\hlstd{,} \hlkwc{nIters} \hlstd{=} \hlnum{10000}\hlstd{,} \hlkwc{seed} \hlstd{=} \hlnum{13} \hlstd{)}
\end{alltt}
\end{kframe}
\end{knitrout}

The output of the function is also a list with the same names as the \code{params} list. Suppose a given parameter has shape $(d_1, \dots, d_l)$, then the output will be an array of shape $(\text{\code{nIters}}, d_1, \dots, d_l)$. So in our case, \code{output\$beta[i,{},]} is the $i^{th}$ MCMC sample from the parameter $\beta$; and \code{dim(output\$beta)} is \code{c(10000, 54, 1)}.

In order to run a control variate algorithm such as \code{sgldcv} we need one additional argument, which is the stepsize for the initial optimisation step. The optimisation uses the \pkg{TensorFlow} \code{GradientDescentOptimizer}. The stepsize should be quite similar to the stepsize for SGLD, though is often slightly larger. First, so that we can measure the performance of the chain, we shall set a seed and randomly remove some observations from the full \code{dataset} to form a \code{testset}. We also remove a short burn-in period of 1000 from the final output.
\begin{knitrout}
\definecolor{shadecolor}{rgb}{0.969, 0.969, 0.969}\color{fgcolor}\begin{kframe}
\begin{alltt}
\hlkwd{set.seed}\hlstd{(}\hlnum{13}\hlstd{)}
\hlstd{testInd} \hlkwb{=} \hlkwd{sample}\hlstd{(}\hlkwd{nrow}\hlstd{(dataset}\hlopt{$}\hlstd{X),} \hlnum{10}\hlopt{^}\hlnum{4}\hlstd{)}
\hlstd{testset} \hlkwb{=} \hlkwd{list}\hlstd{(} \hlstr{"X"} \hlstd{= dataset}\hlopt{$}\hlstd{X[testInd,],} \hlstr{"y"} \hlstd{= dataset}\hlopt{$}\hlstd{y[testInd] )}
\hlstd{dataset} \hlkwb{=} \hlkwd{list}\hlstd{(} \hlstr{"X"} \hlstd{= dataset}\hlopt{$}\hlstd{X[}\hlopt{-}\hlstd{testInd,],} \hlstr{"y"} \hlstd{= dataset}\hlopt{$}\hlstd{y[}\hlopt{-}\hlstd{testInd] )}
\hlstd{output} \hlkwb{=} \hlkwd{sgldcv}\hlstd{( logLik, dataset, params,} \hlnum{5e-6}\hlstd{,} \hlnum{5e-6}\hlstd{,}
  \hlkwc{logPrior} \hlstd{= logPrior,} \hlkwc{minibatchSize} \hlstd{=} \hlnum{500}\hlstd{,} \hlkwc{nIters} \hlstd{=} \hlnum{11000}\hlstd{,} \hlkwc{seed} \hlstd{=} \hlnum{13} \hlstd{)}
\hlstd{output}\hlopt{$}\hlstd{beta} \hlkwb{=} \hlstd{output}\hlopt{$}\hlstd{beta[}\hlopt{-}\hlkwd{c}\hlstd{(}\hlnum{1}\hlopt{:}\hlnum{1000}\hlstd{),,]}
\hlstd{output}\hlopt{$}\hlstd{bias} \hlkwb{=} \hlstd{output}\hlopt{$}\hlstd{bias[}\hlopt{-}\hlkwd{c}\hlstd{(}\hlnum{1}\hlopt{:}\hlnum{1000}\hlstd{)]}
\end{alltt}
\end{kframe}
\end{knitrout}

\begin{figure}[t]
    \centering
    \includegraphics[width=420px]{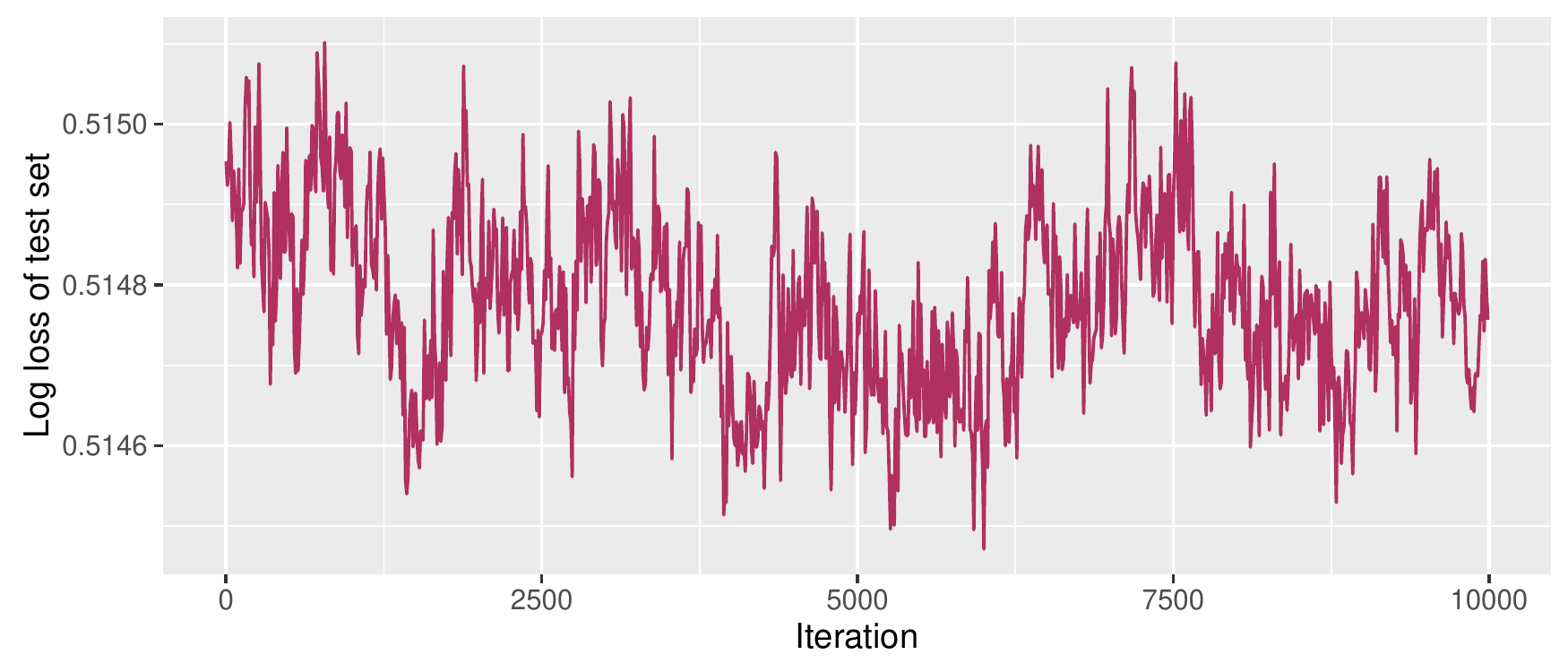}
    \caption{Log loss on a test set for parameters simulated using the \code{sgldcv} algorithm after 1000 iterations of burn-in. Logistic regression problem with the covertype dataset.}
    \label{fig:logReg}
\end{figure}

A common performance measure for a classifier is the log loss. Given an observation with data $\mb x_i$ and response $y_i$, logistic regression predicts that $y_i = 1$ with probability
\[
    \pi(\mb x_i, \beta, \beta_0) = \frac{1}{1 + e^{-\beta_0 - \mb x_i \beta}}.
\]
Given a test set $T$ of data response pairs $(\mb x, y)$, the log loss $s(\cdot)$, of a binary chain, is defined as
\begin{equation}
    s(\beta, \beta_0, T) =  - \frac{1}{|T|} \sum_{(\mb x, y) \in T} \left[ y \log \pi(\mb x, \beta, \beta_0) + (1 - y) \log ( 1 - \pi( \mb x, \beta, \beta_0 )) \right].
    \label{eq:bin-ll}
\end{equation}
To check convergence of \code{sgldcv} we'll plot the log loss every 10 iterations, using the \code{testset} we removed earlier. Results are given in Figure \ref{fig:logReg}. The plot shows the \code{sgldcv} algorithm converging to stationarity after a short burn-in period. The burn-in period is short due to the initial optimisation step. The following chunk of code will calculate the log loss of the output every 10 iterations, and then plot the result.

\begin{knitrout}
\definecolor{shadecolor}{rgb}{0.969, 0.969, 0.969}\color{fgcolor}\begin{kframe}
\begin{alltt}
\hlstd{iterations} \hlkwb{=} \hlkwd{seq}\hlstd{(}\hlkwc{from} \hlstd{=} \hlnum{1}\hlstd{,} \hlkwc{to} \hlstd{=} \hlnum{10}\hlopt{^}\hlnum{4}\hlstd{,} \hlkwc{by} \hlstd{=} \hlnum{10}\hlstd{)}
\hlstd{logLoss} \hlkwb{=} \hlkwd{rep}\hlstd{(}\hlnum{0}\hlstd{,} \hlkwd{length}\hlstd{(iterations))}
\hlkwa{for} \hlstd{( iter} \hlkwa{in} \hlnum{1}\hlopt{:}\hlkwd{length}\hlstd{(iterations) ) \{}
  \hlstd{j} \hlkwb{=} \hlstd{iterations[iter]}
  \hlstd{beta0_j} \hlkwb{=} \hlstd{output}\hlopt{$}\hlstd{bias[j]}
  \hlstd{beta_j} \hlkwb{=} \hlstd{output}\hlopt{$}\hlstd{beta[j,]}
  \hlkwa{for} \hlstd{( i} \hlkwa{in} \hlnum{1}\hlopt{:}\hlkwd{length}\hlstd{(testset}\hlopt{$}\hlstd{y) ) \{}
    \hlstd{piCurr} \hlkwb{=} \hlnum{1} \hlopt{/} \hlstd{(}\hlnum{1} \hlopt{+} \hlkwd{exp}\hlstd{(}\hlopt{-} \hlstd{beta0_j} \hlopt{-} \hlkwd{sum}\hlstd{(testset}\hlopt{$}\hlstd{X[i,]} \hlopt{*} \hlstd{beta_j)))}
    \hlstd{y_i} \hlkwb{=} \hlstd{testset}\hlopt{$}\hlstd{y[i]}
    \hlstd{logLossCurr} \hlkwb{=} \hlopt{-} \hlstd{(y_i} \hlopt{*} \hlkwd{log}\hlstd{(piCurr)} \hlopt{+} \hlstd{(}\hlnum{1} \hlopt{-} \hlstd{y_i)} \hlopt{*} \hlkwd{log}\hlstd{(}\hlnum{1} \hlopt{-} \hlstd{piCurr))}
    \hlstd{logLoss[iter]} \hlkwb{=} \hlstd{logLoss[iter]} \hlopt{+} \hlnum{1} \hlopt{/} \hlkwd{length}\hlstd{(testset}\hlopt{$}\hlstd{y)} \hlopt{*} \hlstd{logLossCurr}
  \hlstd{\}}
\hlstd{\}}
\hlstd{plotFrame} \hlkwb{=} \hlkwd{data.frame}\hlstd{(}\hlstr{"Iteration"} \hlstd{= iterations,} \hlstr{"logLoss"} \hlstd{= logLoss)}
\hlkwd{ggplot}\hlstd{(plotFrame,} \hlkwd{aes}\hlstd{(}\hlkwc{x} \hlstd{= Iteration,} \hlkwc{y} \hlstd{= logLoss))} \hlopt{+}
  \hlkwd{geom_line}\hlstd{(}\hlkwc{color} \hlstd{=} \hlstr{"maroon"}\hlstd{)} \hlopt{+}
  \hlkwd{ylab}\hlstd{(}\hlstr{"Log loss of test set"}\hlstd{)}
\end{alltt}
\end{kframe}
\end{knitrout}

\subsection{Example usage: Storage constraints}
\label{sec:nnusage}

Often large datasets and high dimensionality go hand in hand. Sometimes the dimensionality is so high that storage of the full MCMC chain in memory becomes an issue. There are a number of ways around this, including: calculating estimates of the desired posterior quantity on the fly; reducing the dimensionality of the chain using a test function; or just periodically saving the chain to the hard disk and starting from scratch. To deal with high storage costs, \pkg{sgmcmc} provides functionality to run SGMCMC algorithms step by step. This allows users to deal with the output as they see fit at each iteration.

In this section, we detail how to run SGMCMC chains step by step. To do this requires more knowledge of \pkg{TensorFlow}, including using \pkg{TensorFlow} sessions and building custom placeholders and tensors. For more details see the \pkg{TensorFlow} for \proglang{R} documentation \citep{Tensor4R}. The step by step procedure works similarly to a standard \pkg{TensorFlow} procedure: \pkg{TensorFlow} variables, tensors and placeholders are declared; then the \pkg{TensorFlow} session is started and all the required tensors are initialised; finally the SGMCMC algorithm is run step by step in a user defined loop, and the user evaluates tensors as required.

To demonstrate this concept we keep things simple and use the logistic regression example introduced in the previous section. While this example can fit into memory, it allows us to demonstrate the concepts without getting bogged down in a complicated model. For a more realistic example, where the output does not fit into memory, see the Bayesian neural network model in Section \ref{sec:nn}.

We start by assuming the \code{dataset}, \code{params}, \code{logLik}, \code{logPrior} and \code{stepsize} objects have been created in exactly the same way as in the previous example (Section \ref{sec:usage}). Now suppose we want to make inference using stochastic gradient Langevin dynamics (SGLD), but we want to run it step by step. The first step is to initialise an \code{sgld} object using the function \code{sgldSetup}. For every function in Table \ref{tab:main-fns} there is a corresponding \code{*Setup} function, such as \code{sghmccvSetup} or \code{sgnhtSetup}. This function will create all the \pkg{TensorFlow} objects required, as well as declare the dynamics of the SGMCMC algorithm. For our example we can run the following 

\begin{knitrout}
\definecolor{shadecolor}{rgb}{0.969, 0.969, 0.969}\color{fgcolor}\begin{kframe}
\begin{alltt}
\hlstd{sgld} \hlkwb{=} \hlkwd{sgldSetup}\hlstd{(logLik, dataset, params, stepsize,} \hlkwc{logPrior} \hlstd{= logPrior,}
  \hlkwc{minibatchSize} \hlstd{=} \hlnum{500}\hlstd{,} \hlkwc{seed} \hlstd{=} \hlnum{13}\hlstd{)}
\end{alltt}
\end{kframe}
\end{knitrout}

This \code{sgld} object is a type of \code{sgmcmc} object, it is an \proglang{R} \code{S3} object, which is essentially a list with a number of entries. The most important of these entries for building SGMCMC algorithms is called \code{params}, which holds a list, with the same names as in the \code{params} that were fed to \code{sgldSetup}, but this list contains \code{tf\$Variable} objects. This is how the tensors are accessed which hold the current parameter values in the chain. For more details on the attributes of these objects, see the documentation for \code{sgldSetup}, \code{sgldcvSetup}, etc.

Now that we have created the \code{sgld} object, we want to initialise the \pkg{TensorFlow} variables and the \code{sgmcmc} algorithm chosen. For a standard algorithm, this will initialise the \pkg{TensorFlow} graph and all the tensors that were created. For an algorithm with control variates (e.g., \code{sgldcv}), this will also find the $\hat \theta$ estimates of the parameters and calculate the full log posterior gradient at that point; as detailed in Section \ref{sec:sgmcmccv}. The function used to do this is \code{initSess},

\begin{knitrout}
\definecolor{shadecolor}{rgb}{0.969, 0.969, 0.969}\color{fgcolor}\begin{kframe}
\begin{alltt}
\hlstd{sess} \hlkwb{=} \hlkwd{initSess}\hlstd{(sgld)}
\end{alltt}
\end{kframe}
\end{knitrout}

The \code{sess} returned by \code{initSess} is the current \pkg{TensorFlow} session, which is needed to run the SGMCMC algorithm of choice, and to access any of the tensors needed, such as \code{sgld\$params}.

Now we have everything to run an SGLD algorithm step by step as follows

\begin{knitrout}
\definecolor{shadecolor}{rgb}{0.969, 0.969, 0.969}\color{fgcolor}\begin{kframe}
\begin{alltt}
\hlkwa{for} \hlstd{(i} \hlkwa{in} \hlnum{1}\hlopt{:}\hlnum{10}\hlopt{^}\hlnum{4}\hlstd{) \{}
  \hlkwd{sgmcmcStep}\hlstd{(sgld, sess)}
  \hlstd{currentState} \hlkwb{=} \hlkwd{getParams}\hlstd{(sgld, sess)}
\hlstd{\}}
\end{alltt}
\end{kframe}
\end{knitrout}

Here the function \code{sgmcmcStep} will update \code{sgld\$params} using a single update of SGLD, or whichever SGMCMC algorithm is given. The function \code{getParams} will return a list of the current parameters as \proglang{R} objects rather than as tensors. 

This simple example of running SGLD step by step only stores the most recent value in the chain, which is useless for a Monte Carlo method. Also, for large scale examples, it is often useful to reduce the dimension of the chain by calculating some test function $g(\cdot)$ of $\theta$ at each iteration rather than the parameters themselves. This example will demonstrate how to store a test function at each iteration, as well as calculating the estimated posterior mean on the fly. We assume that a new \proglang{R} session has been started and the \code{sgld} object has just been created using \code{sgldSetup} with the same properties as in the example above. We assume that no \pkg{TensorFlow} session has been created (i.e., \code{initSess} has not been run yet).

Before the \pkg{TensorFlow} session has been declared, the user is able to create their own custom tensors. This is useful, as test functions can be declared beforehand using the \code{sgld\$params} variables, which allows the test functions to be quickly calculated by just evaluating the tensors in the current session. The test function used here is once again the log loss of a test set, as defined in \eqref{eq:bin-ll}.

Suppose we input \code{sgld\$params} and the \code{testset} $T$ to the \code{logLik} function. Then the log loss is actually $- \frac{1}{|T|}$ times this value. This means we can easily create a tensor that calculates the log loss by creating a list of placeholders that hold the test set, then using the \code{logLik} function with the \code{testset} list and \code{sgld\$params} as input. We can do this as follows

\begin{knitrout}
\definecolor{shadecolor}{rgb}{0.969, 0.969, 0.969}\color{fgcolor}\begin{kframe}
\begin{alltt}
\hlstd{testPlaceholder} \hlkwb{=} \hlkwd{list}\hlstd{()}
\hlstd{testPlaceholder[[}\hlstr{"X"}\hlstd{]]} \hlkwb{=} \hlstd{tf}\hlopt{$}\hlkwd{placeholder}\hlstd{(tf}\hlopt{$}\hlstd{float32,} \hlkwd{dim}\hlstd{(testset[[}\hlstr{"X"}\hlstd{]]))}
\hlstd{testPlaceholder[[}\hlstr{"y"}\hlstd{]]} \hlkwb{=} \hlstd{tf}\hlopt{$}\hlkwd{placeholder}\hlstd{(tf}\hlopt{$}\hlstd{float32,} \hlkwd{dim}\hlstd{(testset[[}\hlstr{"y"}\hlstd{]]))}
\hlstd{testSize} \hlkwb{=} \hlkwd{as.double}\hlstd{(}\hlkwd{nrow}\hlstd{(testset[[}\hlstr{"X"}\hlstd{]]))}
\hlstd{logLoss} \hlkwb{=} \hlopt{-} \hlkwd{logLik}\hlstd{(sgld}\hlopt{$}\hlstd{params, testPlaceholder)} \hlopt{/} \hlstd{testSize}
\end{alltt}
\end{kframe}
\end{knitrout}

This placeholder is then fed the full \code{testset} each time the log loss is calculated. Now we will declare the \pkg{TensorFlow} session, and run the chain step by step. At each iteration we will calculate the current Monte Carlo estimate of the parameters. The log loss will be stored every 100 iterations. We omit a plot of the log loss as it is similar to Figure \ref{fig:logReg}. This first chunk of code corresponds to the burn-in, at this point we do not store parameter estimates, we just print the log loss every 100 iterations to check convergence of the chain.
\begin{knitrout}
\definecolor{shadecolor}{rgb}{0.969, 0.969, 0.969}\color{fgcolor}\begin{kframe}
\begin{alltt}
\hlstd{sess} \hlkwb{=} \hlkwd{initSess}\hlstd{(sgld)}
\hlstd{feedDict} \hlkwb{=} \hlkwd{dict}\hlstd{()}
\hlstd{feedDict[[testPlaceholder[[}\hlstr{"X"}\hlstd{]]]]} \hlkwb{=} \hlstd{testset[[}\hlstr{"X"}\hlstd{]]}
\hlstd{feedDict[[testPlaceholder[[}\hlstr{"y"}\hlstd{]]]]} \hlkwb{=} \hlstd{testset[[}\hlstr{"y"}\hlstd{]]}
\hlkwd{message}\hlstd{(}\hlstr{"Burning-in chain..."}\hlstd{)}
\hlkwd{message}\hlstd{(}\hlstr{"iteration\textbackslash{}tlog loss"}\hlstd{)}
\hlkwa{for} \hlstd{(i} \hlkwa{in} \hlnum{1}\hlopt{:}\hlnum{10}\hlopt{^}\hlnum{4}\hlstd{) \{}
  \hlkwa{if} \hlstd{(i} \hlopt{%%} \hlnum{100} \hlopt{==} \hlnum{0}\hlstd{) \{}
    \hlstd{progress} \hlkwb{=} \hlstd{sess}\hlopt{$}\hlkwd{run}\hlstd{(logLoss,} \hlkwc{feed_dict} \hlstd{= feedDict)}
    \hlkwd{message}\hlstd{(}\hlkwd{paste0}\hlstd{(i,} \hlstr{"\textbackslash{}t"}\hlstd{, progress))}
  \hlstd{\}}
  \hlkwd{sgmcmcStep}\hlstd{(sgld, sess)}
\hlstd{\}}
\end{alltt}
\end{kframe}
\end{knitrout}
In the previous chunk of code, the purpose of \code{feedDict} is to feed the whole \code{testset} to the \code{logLoss} tensor. In the next chunk of code we run the chain for $10^4$ iterations. At each iteration we store a running average of the parameter estimates in the \proglang{R} list object \code{postMean}. Again we print the log loss of the chain every 100 iterations.
\begin{knitrout}
\definecolor{shadecolor}{rgb}{0.969, 0.969, 0.969}\color{fgcolor}\begin{kframe}
\begin{alltt}
\hlstd{postMean} \hlkwb{=} \hlkwd{getParams}\hlstd{(sgld, sess)}
\hlstd{logLossOut} \hlkwb{=} \hlkwd{rep}\hlstd{(}\hlnum{0}\hlstd{,} \hlnum{10}\hlopt{^}\hlnum{4} \hlopt{/} \hlnum{100}\hlstd{)}
\hlkwd{message}\hlstd{(}\hlstr{"Running SGMCMC..."}\hlstd{)}
\hlkwa{for} \hlstd{(i} \hlkwa{in} \hlnum{1}\hlopt{:}\hlnum{10}\hlopt{^}\hlnum{4}\hlstd{) \{}
  \hlkwd{sgmcmcStep}\hlstd{(sgld, sess)}
  \hlstd{currentState} \hlkwb{=} \hlkwd{getParams}\hlstd{(sgld, sess)}
  \hlkwa{for} \hlstd{(paramName} \hlkwa{in} \hlkwd{names}\hlstd{(postMean)) \{}
    \hlstd{postMean[[paramName]]} \hlkwb{=} \hlstd{(postMean[[paramName]]} \hlopt{*} \hlstd{i} \hlopt{+}
      \hlstd{currentState[[paramName]])} \hlopt{/} \hlstd{(i} \hlopt{+} \hlnum{1}\hlstd{)}
  \hlstd{\}}
  \hlkwa{if} \hlstd{(i} \hlopt{%%} \hlnum{100} \hlopt{==} \hlnum{0}\hlstd{) \{}
    \hlstd{logLossOut[i}\hlopt{/}\hlnum{100}\hlstd{]} \hlkwb{=} \hlstd{sess}\hlopt{$}\hlkwd{run}\hlstd{(logLoss,} \hlkwc{feed_dict} \hlstd{= feedDict)}
    \hlkwd{message}\hlstd{(}\hlkwd{paste0}\hlstd{(i,} \hlstr{"\textbackslash{}t"}\hlstd{, logLossOut[i}\hlopt{/}\hlnum{100}\hlstd{]))}
  \hlstd{\}}
\hlstd{\}}
\end{alltt}
\end{kframe}
\end{knitrout}

\section{Simulations}
\label{sec:simulations}

In this section we demonstrate the algorithms and performance by simulating from a variety of models using all the implemented methods and commenting on the performance of each. These simulations are reproducible and available in the supplementary material and on Github.\footnote{\url{https://github.com/jbaker92/sgmcmc-simulations}} For more usage tutorials similar to Sections \ref{sec:usage} and \ref{sec:nnusage}, please see the vignettes on the package website.\footnote{\url{https://stor-i.github.io/sgmcmc}}

\subsection{Gaussian mixture}
\label{sec:sim-gm}

In this model we assume our dataset $x_1, \dots, x_N$ is drawn i.i.d from
\begin{equation}
    X_i \, | \, \theta_1, \theta_2 \sim \frac{1}{2} \mathcal{N}(\theta_1, \mathrm I_2) + \frac{1}{2} \mathcal{N}(\theta_2, \mathrm I_2), \quad i = 1, \dots, N; 
    \label{eq:sim-gm}
\end{equation}
where $\theta_1, \theta_2$ are parameters to be inferred and $\mathrm I_2$ is the $2 \times 2$ identity matrix. We assume the prior $\theta_1, \theta_2 \sim \mathcal{N}(0, 10 \mathrm I_2$). To generate the synthetic dataset, we simulate $10^3$ observations from $\frac{1}{2} \mathcal{N}\left([0, 0]^\top, \mathrm I_2\right) + \frac{1}{2} \mathcal{N}\left([0.1, 0.1]^\top, \mathrm I_2\right)$. While this is a small number of observations, it allows us to compare the results to a full HMC scheme using the \proglang{R} implementation of \pkg{Stan} \citep{Carpenter2016}. The full HMC scheme should sample from close to the true posterior distribution, so acts as a good surrogate for the truth. We compare each \pkg{sgmcmc} algorithm implemented to the HMC sample to compare performance. Larger scale examples are given in Sections \ref{sec:sim-lr} and \ref{sec:nn}. We ran all methods for $10^4$ iterations, except SGHMC, since the computational cost is greater for this method due to the trajectory parameter $L$. We ran SGHMC for 2,000 iterations, using default trajectory $L = 5$, as this ensures the overall computational cost of the method is similar to the other methods. We used a burn-in step of $10^4$ iterations, except for the control variate methods, where we used $10^4$ iterations in the initial optimisation step, with no burn-in. Again this ensures comparable computational cost across different methods.

The \code{logLik} and \code{logPrior} functions used for this model are as follows. Remember that any distribution can be referenced using \code{tf\$distributions} once the package has been loaded in.
\begin{knitrout}
\definecolor{shadecolor}{rgb}{0.969, 0.969, 0.969}\color{fgcolor}\begin{kframe}
\begin{alltt}
\hlstd{logLik} \hlkwb{=} \hlkwa{function}\hlstd{(} \hlkwc{params}\hlstd{,} \hlkwc{dataset} \hlstd{) \{}
  \hlstd{SigmaDiag} \hlkwb{=} \hlkwd{c}\hlstd{(}\hlnum{1}\hlstd{,} \hlnum{1}\hlstd{)}
  \hlstd{component1} \hlkwb{=} \hlstd{tf}\hlopt{$}\hlstd{distributions}\hlopt{$}\hlkwd{MultivariateNormalDiag}\hlstd{(}
    \hlstd{params}\hlopt{$}\hlstd{theta1, SigmaDiag )}
  \hlstd{component2} \hlkwb{=} \hlstd{tf}\hlopt{$}\hlstd{distributions}\hlopt{$}\hlkwd{MultivariateNormalDiag}\hlstd{(}
    \hlstd{params}\hlopt{$}\hlstd{theta2, SigmaDiag )}
  \hlstd{probs} \hlkwb{=} \hlstd{tf}\hlopt{$}\hlstd{distributions}\hlopt{$}\hlkwd{Categorical}\hlstd{(}\hlkwd{c}\hlstd{(}\hlnum{0.5}\hlstd{,}\hlnum{0.5}\hlstd{))}
  \hlstd{distn} \hlkwb{=} \hlstd{tf}\hlopt{$}\hlstd{distributions}\hlopt{$}\hlkwd{Mixture}\hlstd{(}
    \hlstd{probs,} \hlkwd{list}\hlstd{(component1, component2))}
  \hlstd{logLik} \hlkwb{=} \hlstd{tf}\hlopt{$}\hlkwd{reduce_sum}\hlstd{( distn}\hlopt{$}\hlkwd{log_prob}\hlstd{(dataset}\hlopt{$}\hlstd{X) )}
  \hlkwd{return}\hlstd{( logLik )}
\hlstd{\}}

\hlstd{logPrior} \hlkwb{=} \hlkwa{function}\hlstd{(} \hlkwc{params} \hlstd{) \{}
  \hlstd{mu0} \hlkwb{=} \hlkwd{c}\hlstd{(} \hlnum{0}\hlstd{,} \hlnum{0} \hlstd{)}
  \hlstd{Sigma0Diag} \hlkwb{=} \hlkwd{c}\hlstd{(}\hlnum{10}\hlstd{,} \hlnum{10}\hlstd{)}
  \hlstd{priorDistn} \hlkwb{=} \hlstd{tf}\hlopt{$}\hlstd{distributions}\hlopt{$}\hlkwd{MultivariateNormalDiag}\hlstd{(}
    \hlstd{mu0, Sigma0Diag )}
  \hlstd{logPrior} \hlkwb{=} \hlstd{priorDistn}\hlopt{$}\hlkwd{log_prob}\hlstd{( params}\hlopt{$}\hlstd{theta1 )} \hlopt{+}
    \hlstd{priorDistn}\hlopt{$}\hlkwd{log_prob}\hlstd{( params}\hlopt{$}\hlstd{theta2 )}
  \hlkwd{return}\hlstd{( logPrior )}
\hlstd{\}}
\end{alltt}
\end{kframe}
\end{knitrout}

The following list determines the stepsizes used for each method, the \code{optStepsize} used for control variate methods was \code{5e-5}.
\begin{knitrout}
\definecolor{shadecolor}{rgb}{0.969, 0.969, 0.969}\color{fgcolor}\begin{kframe}
\begin{alltt}
\hlstd{stepsizeList} \hlkwb{=} \hlkwd{list}\hlstd{(}\hlstr{"sgld"} \hlstd{=} \hlnum{5e-3}\hlstd{,} \hlstr{"sghmc"} \hlstd{=} \hlnum{5e-4}\hlstd{,} \hlstr{"sgnht"} \hlstd{=} \hlnum{3e-4}\hlstd{,}
  \hlstr{"sgldcv"} \hlstd{=} \hlnum{1e-2}\hlstd{,} \hlstr{"sghmccv"} \hlstd{=} \hlnum{1.5e-3}\hlstd{,} \hlstr{"sgnhtcv"} \hlstd{=} \hlnum{3e-3}\hlstd{)}
\end{alltt}
\end{kframe}
\end{knitrout}

We set the seed to be 2 using the optional \code{seed} argument and use a minibatch size of 100. We also used a seed of 2 when generating the data (see the supplementary material for full details). Starting points were sampled from a standard normal.

\begin{figure}[t]
    \centering
    \includegraphics[width=400px]{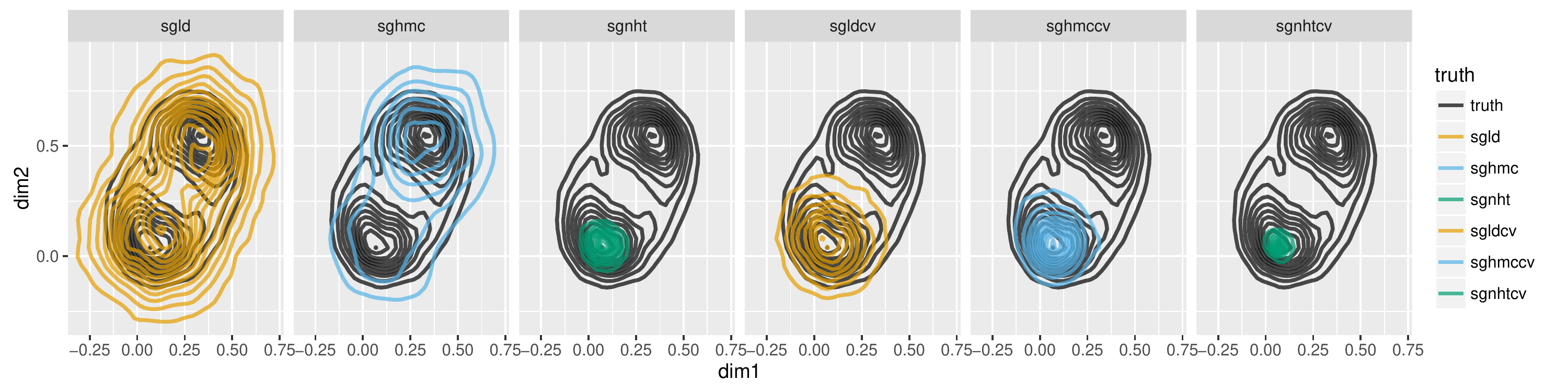}
    \caption{Plots of the approximate posterior for $\theta_1$ simulated using each of the methods implemented by \pkg{sgmcmc}, compared with a full HMC run, treated as the truth, for the Gaussian mixture model \eqref{eq:sim-gm}.}
    \label{fig:sim-gm}
\end{figure}

The results are plotted in Figure \ref{fig:sim-gm}. The black contours represent the best guess at the true posterior, which was found using the standard HMC procedure in \pkg{Stan}. The coloured contours that overlay the black contours are the approximations of each of the SGMCMC methods implemented by \pkg{sgmcmc}. This allows us to compare the SGMCMC estimates with the `truth' by eye.

In the simulation, we obtain two chains, one approximating $\theta_1$ and the other approximating $\theta_2$. In order to examine how well the methods explore both modes, we take just $\theta_1$ and compare this to the HMC run for $\theta_1$. The results are quite variable, and it demonstrates a point nicely: there seems to be a trade-off between predictive accuracy and exploration. Many methods have demonstrated good performance using predictive accuracy; where a test set is removed from the full dataset to assess how well the fitted model performs on the test set. This is a useful technique for complex models, which are high dimensional and have a large number of data points, as they cannot be plotted, and an MCMC run to act as the `truth' cannot be fitted.

However, this example shows that it does not give the full picture. A lot of the methods which show improved predictive performance (e.g., control variate methods and especially \code{sgnht}) over \code{sgld} appear here to perform worse at exploring the full space. In this example, \code{sgld} performs the best at exploring both components, though it over-estimates posterior uncertainty. The algorithm \code{sghmc} also explores both components but somewhat unevenly. We find that \code{sgnht}, while being shown to have better predictive performance in the original paper \citep{Ding2014}, does not do nearly as well as the other algorithms at exploring the space and appears to collapse to the posterior mode. The control variate methods, shown in the following sections, and in \cite{Baker2017}, appear to have better predictive performance than \code{sgld}, but do not explore both components either. For example, \code{sgldcv} explores the space the best but over-estimates uncertainty of the first component, since it relies on SGLD updates which also overestimates uncertainty. In contrast, \code{sgnhtcv} collapses to a posterior mode since it relies on the SGNHT updates which also collapse.

\subsection{Bayesian logistic regression}
\label{sec:sim-lr}

In this section, we apply all the methods to the logistic regression example in Section \ref{sec:usage}. We compare the performance of the methods by calculating the log loss of a test set every 10 iterations, again as detailed in Section \ref{sec:usage}. The standard methods (\code{sgld}, \code{sghmc}, \code{sgnht}) were run for $10^4$ iterations with an additional $10^4$ iterations of burn-in; except for \code{sghmc} which has $5 \times$ the computational cost, so is ran for 2,000 iterations with 2,000 iterations of burn-in. The control variate methods were run for $10^4$ iterations with an additional $10^4$ iterations for the initial optimisation step, and no burn-in; again except for \code{sghmccv} which was run for 2,000 iterations. This means that all the methods should be somewhat comparable in terms of computation time.

\begin{figure}[t]
    \centering
    \includegraphics[width=400px]{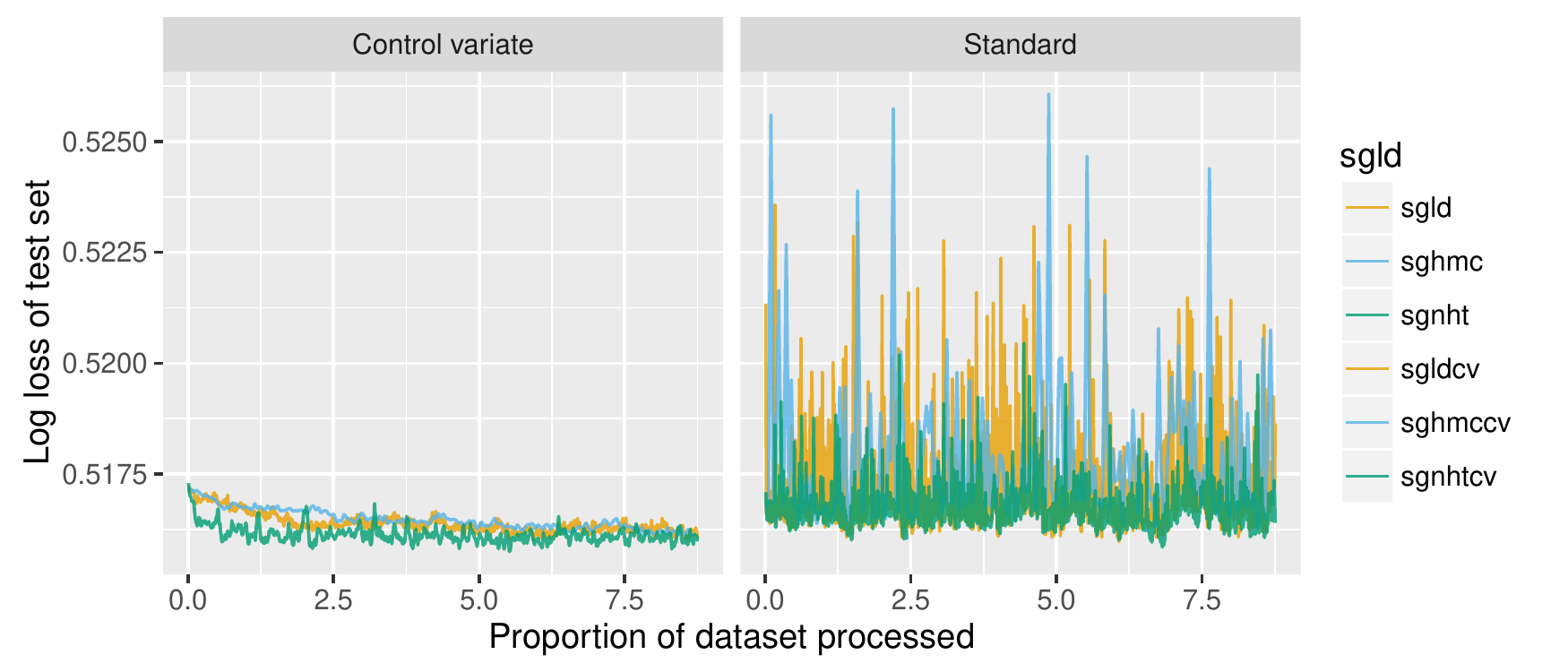}
    \caption{Plots of the log loss of a test set for $\beta_0$ and $\beta$ simulated using each of the methods implemented by \pkg{sgmcmc}. Logistic regression problem with the covertype dataset.}
    \label{fig:sim-lr}
\end{figure}

The following list determines the stepsizes used for each method, the \code{optStepsize} used was \code{1e-6}.
\begin{knitrout}
\definecolor{shadecolor}{rgb}{0.969, 0.969, 0.969}\color{fgcolor}\begin{kframe}
\begin{alltt}
\hlstd{stepsizes} \hlkwb{=} \hlkwd{list}\hlstd{(}\hlstr{"sgld"} \hlstd{=} \hlnum{5e-6}\hlstd{,} \hlstr{"sghmc"} \hlstd{=} \hlnum{1e-7}\hlstd{,} \hlstr{"sgnht"} \hlstd{=} \hlnum{1e-7}\hlstd{,} \hlstr{"sgldcv"} \hlstd{=} \hlnum{1e-5}\hlstd{,}
  \hlstr{"sghmccv"} \hlstd{=} \hlnum{1e-6}\hlstd{,} \hlstr{"sgnhtcv"} \hlstd{=} \hlnum{5e-7}\hlstd{)}
\end{alltt}
\end{kframe}
\end{knitrout}
We set the seed to be 1 for each of the simulations, and when generating the test data (see the supplementary material for reproducible code) and use a minibatch size of 500. Starting points are sampled from a standard normal.

Results are plotted in Figure \ref{fig:sim-lr}. All of the algorithms show decent performance. Methods which use control variates have significantly better predictive performance; and result in chains with lower variance. \code{sghmc} has lower variance than \code{sgld} and \code{sgnht}, though this could be related to the high computational cost. One might envisage setting a lower trajectory $L$ would result in a chain with higher variance. \code{sgldcv} takes longer to burn-in  than the other control variate methods. The algorithm \code{sgld} has the highest variance by far; this could be related to our discussion in Section \ref{sec:sim-gm} on exploration versus accuracy.

\subsection{Bayesian neural network}
\label{sec:nn}

In this simulation we demonstrate a very high dimensional chain. This gives a more realistic example of when we would want to run the chain step by step. The model is a two layer Bayesian neural network which is fit to the MNIST dataset \citep{Lecun2010}. More step by step detail of how to fit this model can be found in the \pkg{sgmcmc} vignettes or on the package website\footnote{\url{https://stor-i.github.io/sgmcmc/articles/nn.html}}. The MNIST dataset consists of $28 \times 28$ pixel images of handwritten digits from zero to nine. The images are flattened to be a vector of length 784. The dataset is available as a standard dataset from the \pkg{TensorFlow} library, with a matrix of 55,000 training vectors and 10,000 test vectors, each with their corresponding labels. The dataset can be constructed in a similar way to the logistic regression example of Section \ref{sec:usage}, using the standard dataset in the package \code{mnist}.
\begin{knitrout}
\definecolor{shadecolor}{rgb}{0.969, 0.969, 0.969}\color{fgcolor}\begin{kframe}
\begin{alltt}
\hlkwd{library}\hlstd{(}\hlstr{"sgmcmc"}\hlstd{)}
\hlstd{mnist} \hlkwb{=} \hlkwd{getDataset}\hlstd{(}\hlstr{"mnist"}\hlstd{)}
\hlstd{dataset} \hlkwb{=} \hlkwd{list}\hlstd{(}\hlstr{"X"} \hlstd{= mnist}\hlopt{$}\hlstd{train}\hlopt{$}\hlstd{images,} \hlstr{"y"} \hlstd{= mnist}\hlopt{$}\hlstd{train}\hlopt{$}\hlstd{labels)}
\hlstd{testset} \hlkwb{=} \hlkwd{list}\hlstd{(}\hlstr{"X"} \hlstd{= mnist}\hlopt{$}\hlstd{test}\hlopt{$}\hlstd{images,} \hlstr{"y"} \hlstd{= mnist}\hlopt{$}\hlstd{test}\hlopt{$}\hlstd{labels)}
\end{alltt}
\end{kframe}
\end{knitrout}

We build the same neural network model as in the original SGHMC paper by \cite{Chen2014}. Suppose $Y_i$ takes values in $\{0,\dots,9\}$, so is the output label of a digit, and $\mathbf x_i$ is the input vector, with $\mathbf X$ the full $N \times 784$ dataset, where $N$ is the number of observations. The model is then as follows
\begin{align}
    Y_i \, | \, \theta, \mathbf x_i \sim \text{Categorical}( \beta(\theta, \mathbf x_i) ), \\
    \beta(\theta, \mathbf x_i) = \sigma \left( \sigma \left( \mathbf x_i^\top B + b \right) A + a \right).
    \label{eq:beta}
\end{align}
Here $A$, $B$, $a$, $b$ are parameters to be inferred with $\theta = (A, B, a, b)$; $\sigma(\cdot)$ is the softmax function (a generalisation of the logistic link function). $A$, $B$, $a$ and $b$ are matrices with dimensions: $100 \times 10$, $784 \times 100$, $1 \times 10$ and $1 \times 100$ respectively. Each element of these parameters is assigned a normal prior 
\begin{align*}
    A_{kl} | \lambda_A \sim \mathcal{N}(0, \lambda_A^{-1}), \quad B_{jk} | \lambda_B \sim \mathcal{N}(0, \lambda_B^{-1}), \\
    a_l | \lambda_a \sim \mathcal{N}(0, \lambda_a^{-1}), \quad b_k | \lambda_b \sim \mathcal{N}(0, \lambda_b^{-1}), \\
    j = 1,\dots,784; \quad k = 1,\dots,100; \quad l = 1,\dots,10;
\end{align*}
where $\lambda_A$, $\lambda_B$, $\lambda_a$ and $\lambda_b$ are hyperparameters. Finally, we assume
\[
    \lambda_A, \lambda_B, \lambda_a, \lambda_b \sim \text{Gamma}(1, 1).
\]

The model contains a large number of high dimensional parameters, and unless there is sufficient RAM available, a standard chain of length $10^4$ will not fit into memory. First, we shall create the \code{params} dictionary, and then code the \code{logLik} and \code{logPrior} functions. We can sample the initial $\lambda$ parameters from a standard Gamma distribution, and the remaining parameters from a standard normal as follows

\begin{knitrout}
\definecolor{shadecolor}{rgb}{0.969, 0.969, 0.969}\color{fgcolor}\begin{kframe}
\begin{alltt}
\hlstd{d} \hlkwb{=} \hlkwd{ncol}\hlstd{(dataset}\hlopt{$}\hlstd{X)}
\hlstd{params} \hlkwb{=} \hlkwd{list}\hlstd{()}
\hlstd{params}\hlopt{$}\hlstd{A} \hlkwb{=} \hlkwd{matrix}\hlstd{(}\hlkwd{rnorm}\hlstd{(}\hlnum{10}\hlopt{*}\hlnum{100}\hlstd{),} \hlkwc{ncol} \hlstd{=} \hlnum{10}\hlstd{)}
\hlstd{params}\hlopt{$}\hlstd{B} \hlkwb{=} \hlkwd{matrix}\hlstd{(}\hlkwd{rnorm}\hlstd{(d}\hlopt{*}\hlnum{100}\hlstd{),} \hlkwc{ncol} \hlstd{=} \hlnum{100}\hlstd{)}
\hlstd{params}\hlopt{$}\hlstd{a} \hlkwb{=} \hlkwd{rnorm}\hlstd{(}\hlnum{10}\hlstd{)}
\hlstd{params}\hlopt{$}\hlstd{b} \hlkwb{=} \hlkwd{rnorm}\hlstd{(}\hlnum{100}\hlstd{)}
\hlstd{params}\hlopt{$}\hlstd{lambdaA} \hlkwb{=} \hlkwd{rgamma}\hlstd{(}\hlnum{1}\hlstd{,} \hlnum{1}\hlstd{)}
\hlstd{params}\hlopt{$}\hlstd{lambdaB} \hlkwb{=} \hlkwd{rgamma}\hlstd{(}\hlnum{1}\hlstd{,} \hlnum{1}\hlstd{)}
\hlstd{params}\hlopt{$}\hlstd{lambdaa} \hlkwb{=} \hlkwd{rgamma}\hlstd{(}\hlnum{1}\hlstd{,} \hlnum{1}\hlstd{)}
\hlstd{params}\hlopt{$}\hlstd{lambdab} \hlkwb{=} \hlkwd{rgamma}\hlstd{(}\hlnum{1}\hlstd{,} \hlnum{1}\hlstd{)}
\end{alltt}
\end{kframe}
\end{knitrout}

Next we can declare the \code{logLik} function, as based on \eqref{eq:beta}, using standard \code{TensorFlow} methods.

\begin{knitrout}
\definecolor{shadecolor}{rgb}{0.969, 0.969, 0.969}\color{fgcolor}\begin{kframe}
\begin{alltt}
\hlstd{logLik} \hlkwb{=} \hlkwa{function}\hlstd{(}\hlkwc{params}\hlstd{,} \hlkwc{dataset}\hlstd{) \{}
  \hlstd{beta} \hlkwb{=} \hlstd{tf}\hlopt{$}\hlstd{nn}\hlopt{$}\hlkwd{softmax}\hlstd{(tf}\hlopt{$}\hlkwd{matmul}\hlstd{(dataset}\hlopt{$}\hlstd{X, params}\hlopt{$}\hlstd{B)} \hlopt{+} \hlstd{params}\hlopt{$}\hlstd{b)}
  \hlstd{beta} \hlkwb{=} \hlstd{tf}\hlopt{$}\hlstd{nn}\hlopt{$}\hlkwd{softmax}\hlstd{(tf}\hlopt{$}\hlkwd{matmul}\hlstd{(beta, params}\hlopt{$}\hlstd{A)} \hlopt{+} \hlstd{params}\hlopt{$}\hlstd{a)}
  \hlstd{logLik} \hlkwb{=} \hlstd{tf}\hlopt{$}\hlkwd{reduce_sum}\hlstd{(dataset}\hlopt{$}\hlstd{y} \hlopt{*} \hlstd{tf}\hlopt{$}\hlkwd{log}\hlstd{(beta))}
  \hlkwd{return}\hlstd{(logLik)}
\hlstd{\}}
\end{alltt}
\end{kframe}
\end{knitrout}

The \code{logPrior} function can be declared as follows. Remember that any distribution can be referenced using \code{tf\$distributions} once the package has been loaded in.

\begin{knitrout}
\definecolor{shadecolor}{rgb}{0.969, 0.969, 0.969}\color{fgcolor}\begin{kframe}
\begin{alltt}
\hlstd{logPrior} \hlkwb{=} \hlkwa{function}\hlstd{(}\hlkwc{params}\hlstd{) \{}
  \hlstd{distLambda} \hlkwb{=} \hlstd{tf}\hlopt{$}\hlstd{distributions}\hlopt{$}\hlkwd{Gamma}\hlstd{(}\hlnum{1}\hlstd{,} \hlnum{1}\hlstd{)}
  \hlstd{distA} \hlkwb{=} \hlstd{tf}\hlopt{$}\hlstd{distributions}\hlopt{$}\hlkwd{Normal}\hlstd{(}\hlnum{0}\hlstd{, tf}\hlopt{$}\hlkwd{rsqrt}\hlstd{(params}\hlopt{$}\hlstd{lambdaA))}
  \hlstd{logPriorA} \hlkwb{=} \hlstd{tf}\hlopt{$}\hlkwd{reduce_sum}\hlstd{(distA}\hlopt{$}\hlkwd{log_prob}\hlstd{(params}\hlopt{$}\hlstd{A))} \hlopt{+}
    \hlstd{distLambda}\hlopt{$}\hlkwd{log_prob}\hlstd{(params}\hlopt{$}\hlstd{lambdaA)}
  \hlstd{distB} \hlkwb{=} \hlstd{tf}\hlopt{$}\hlstd{distributions}\hlopt{$}\hlkwd{Normal}\hlstd{(}\hlnum{0}\hlstd{, tf}\hlopt{$}\hlkwd{rsqrt}\hlstd{(params}\hlopt{$}\hlstd{lambdaB))}
  \hlstd{logPriorB} \hlkwb{=} \hlstd{tf}\hlopt{$}\hlkwd{reduce_sum}\hlstd{(distB}\hlopt{$}\hlkwd{log_prob}\hlstd{(params}\hlopt{$}\hlstd{B))} \hlopt{+}
    \hlstd{distLambda}\hlopt{$}\hlkwd{log_prob}\hlstd{(params}\hlopt{$}\hlstd{lambdaB)}
  \hlstd{dista} \hlkwb{=} \hlstd{tf}\hlopt{$}\hlstd{distributions}\hlopt{$}\hlkwd{Normal}\hlstd{(}\hlnum{0}\hlstd{, tf}\hlopt{$}\hlkwd{rsqrt}\hlstd{(params}\hlopt{$}\hlstd{lambdaa))}
  \hlstd{logPriora} \hlkwb{=} \hlstd{tf}\hlopt{$}\hlkwd{reduce_sum}\hlstd{(dista}\hlopt{$}\hlkwd{log_prob}\hlstd{(params}\hlopt{$}\hlstd{a))} \hlopt{+}
    \hlstd{distLambda}\hlopt{$}\hlkwd{log_prob}\hlstd{(params}\hlopt{$}\hlstd{lambdaa)}
  \hlstd{distb} \hlkwb{=} \hlstd{tf}\hlopt{$}\hlstd{distributions}\hlopt{$}\hlkwd{Normal}\hlstd{(}\hlnum{0}\hlstd{, tf}\hlopt{$}\hlkwd{rsqrt}\hlstd{(params}\hlopt{$}\hlstd{lambdab))}
  \hlstd{logPriorb} \hlkwb{=} \hlstd{tf}\hlopt{$}\hlkwd{reduce_sum}\hlstd{(distb}\hlopt{$}\hlkwd{log_prob}\hlstd{(params}\hlopt{$}\hlstd{b))} \hlopt{+}
    \hlstd{distLambda}\hlopt{$}\hlkwd{log_prob}\hlstd{(params}\hlopt{$}\hlstd{lambdab)}
  \hlstd{logPrior} \hlkwb{=} \hlstd{logPriorA} \hlopt{+} \hlstd{logPriorB} \hlopt{+} \hlstd{logPriora} \hlopt{+} \hlstd{logPriorb}
  \hlkwd{return}\hlstd{(logPrior)}
\hlstd{\}}
\end{alltt}
\end{kframe}
\end{knitrout}

Similar to Section \ref{sec:nnusage}, we use the log loss as a test function. This time though it is necessary to update the definition, as the logistic regression example was a binary problem whereas now we have a multiclass problem. Given a test set $T$ of pairs $(\mb x, y)$, now $y$ can take values in $\{0, \dots, K\}$, rather than just binary values. To account for this we redefine the definition of log loss to be
\[
    s(\theta, T) = - \frac{1}{|T|} \sum_{\mb x, y \in T} \sum_{k=1}^K \mathbf 1_{y = k} \log \beta_k( \theta, \mathbf x ),
\]
where $\mathbf 1_A$ is the indicator function, and $\beta_k(\theta, \mb x)$ is the $k^{th}$ element of $\beta(\theta, \mb x)$ as defined in \eqref{eq:beta}.

As in Section \ref{sec:nnusage}, the log loss is simply $- \frac{1}{|T|}$ times the \code{logLik} function, if we feed it the \code{testset} rather than the \code{dataset}. This means the \code{logLoss} tensor can be declared in a similar way to Section \ref{sec:nnusage}

\begin{knitrout}
\definecolor{shadecolor}{rgb}{0.969, 0.969, 0.969}\color{fgcolor}\begin{kframe}
\begin{alltt}
\hlstd{testPlaceholder} \hlkwb{=} \hlkwd{list}\hlstd{()}
\hlstd{testPlaceholder[[}\hlstr{"X"}\hlstd{]]} \hlkwb{=} \hlstd{tf}\hlopt{$}\hlkwd{placeholder}\hlstd{(tf}\hlopt{$}\hlstd{float32,} \hlkwd{dim}\hlstd{(testset[[}\hlstr{"X"}\hlstd{]]))}
\hlstd{testPlaceholder[[}\hlstr{"y"}\hlstd{]]} \hlkwb{=} \hlstd{tf}\hlopt{$}\hlkwd{placeholder}\hlstd{(tf}\hlopt{$}\hlstd{float32,} \hlkwd{dim}\hlstd{(testset[[}\hlstr{"y"}\hlstd{]]))}
\hlstd{testSize} \hlkwb{=} \hlkwd{as.double}\hlstd{(}\hlkwd{nrow}\hlstd{(testset[[}\hlstr{"X"}\hlstd{]]))}
\hlstd{logLoss} \hlkwb{=} \hlopt{-} \hlkwd{logLik}\hlstd{(sgld}\hlopt{$}\hlstd{params, testPlaceholder)} \hlopt{/} \hlstd{testSize}
\end{alltt}
\end{kframe}
\end{knitrout}

\begin{figure}[t]
    \centering
    \includegraphics[width=400px]{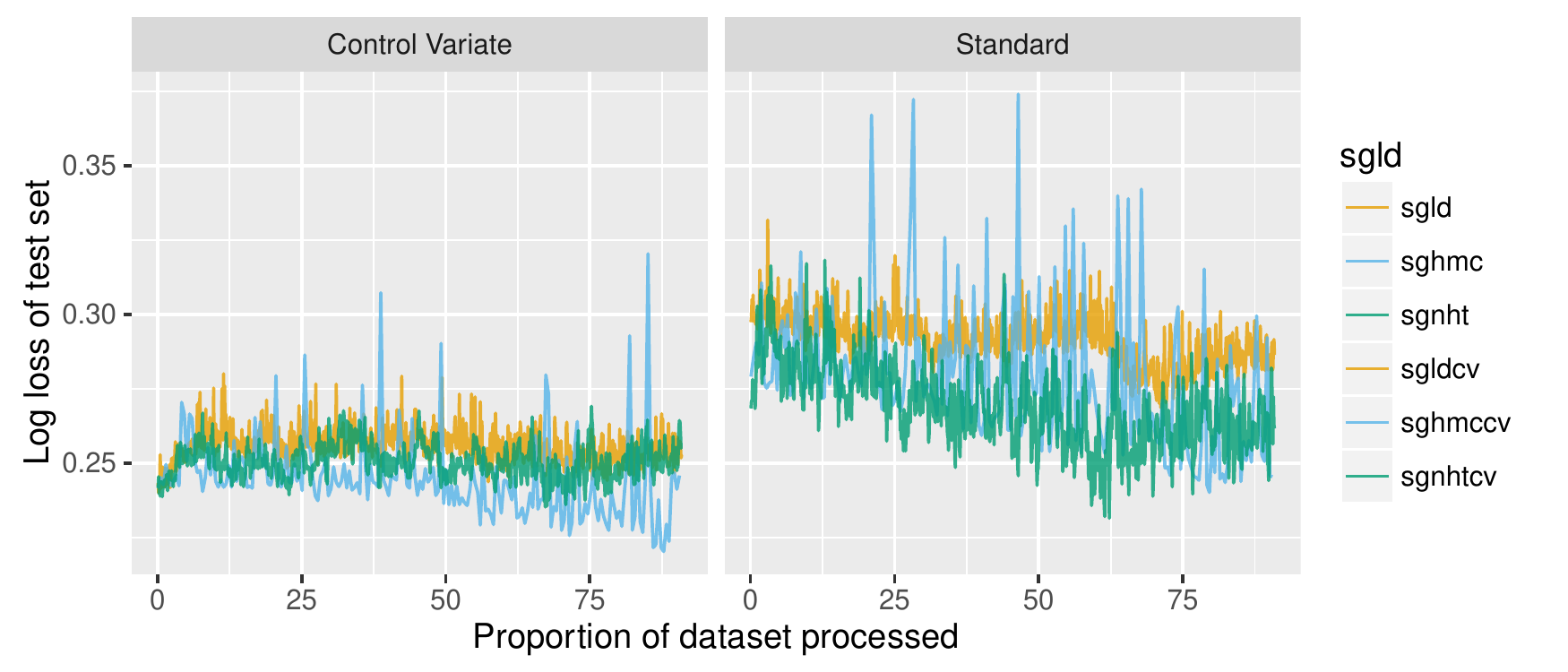}
    \caption{Plots of the log loss of a held out test set based on simulations using each of the methods implemented by \pkg{sgmcmc}. Bayesian neural network problem with MNIST dataset.}
    \label{fig:sim-nn}
\end{figure}

We can run the chain in exactly the same way as Section \ref{sec:nnusage}, and so omit the code for this. We ran $10^4$ iterations of each of the algorithms in Table \ref{tab:main-fns}, calculating the log loss for each algorithm every 10 iterations. The standard algorithms have $10^4$ iterations of burn-in while the control variate algorithms have no burn-in, but $10^4$ iterations in the initial optimisation step. Note that due to the trajectory parameter $L$, \code{sghmc} and \code{sghmccv} will have 5 times the computational cost of the other algorithms. Therefore, we ran these algorithms for 2,000 iterations instead, to make the computational cost comparable. We used the following list of stepsizes
\begin{knitrout}
\definecolor{shadecolor}{rgb}{0.969, 0.969, 0.969}\color{fgcolor}\begin{kframe}
\begin{alltt}
\hlkwd{list}\hlstd{(}\hlstr{"sgld"} \hlstd{=} \hlnum{1e-4}\hlstd{,} \hlstr{"sghmc"} \hlstd{=} \hlnum{1e-5}\hlstd{,} \hlstr{"sgnht"} \hlstd{=} \hlnum{5e-6}\hlstd{,} \hlstr{"sgldcv"} \hlstd{=} \hlnum{5e-5}\hlstd{,}
  \hlstr{"sghmccv"} \hlstd{=} \hlnum{1e-5}\hlstd{,} \hlstr{"sgnhtcv"} \hlstd{=} \hlnum{5e-7}\hlstd{)}
\end{alltt}
\end{kframe}
\end{knitrout}
Generally these are the stepsizes which produce the smallest log loss; except when these chains did not seem to explore the space fully, in which case we increased the stepsize slightly. We set the seed to be 1 for each of the simulations, and when generating the test data (see the supplementary material for reproducible code).

The results are plotted in Figure \ref{fig:sim-nn}. Again we see improvements in the predictive performance of the control variate methods. Among the standard methods, \code{sghmc} and \code{sgnht} have the best predictive performance; which is to be expected given the apparent trade-off between accuracy and exploration.

\section{Discussion}
\label{sec:discussion}

We presented the \proglang{R} package \pkg{sgmcmc}, which enables Bayesian inference with large datasets using stochastic gradient Markov chain Monte Carlo. The package only requires the user to specify the log likelihood and log prior functions; and any differentiation required can be performed automatically. The package is based on \pkg{TensorFlow}, an efficient library for numerical computation that can take advantage of many different architectures, including GPUs. The \pkg{sgmcmc} package keeps much of this efficiency. The package provides functionality to deal with cases where the full MCMC chain is too large to fit into memory. As the chain can be run step by step at each iteration, there is flexibility for these cases.

We implemented the methods on a variety of statistical models, many on realistic datasets. One of these statistical models was a neural network, for which the full MCMC chain would not fit into memory. In this case we demonstrated building test functions and calculating the Monte Carlo estimates on the fly. We empirically demonstrated the predictive performance of the algorithms and the trade-off that appears to occur between predictive performance and exploration.

Many complex models for which SGMCMC methods have been found to perform well require Gibbs updates to be performed periodically \citep{{Patterson2013,Li2016}}. In the future we would like to build functionality for user defined Gibbs steps that can be updated step by step alongside the \pkg{sgmcmc} algorithms. SGHMC has been implemented by setting the value $\hat \beta_t = 0$, as in the experiments of the original paper \cite{Chen2014}. In the future, we would like to implement a more sophisticated approach to set this value, such as using a similar estimate to \cite{Ahn2012}.

\section{Acknowledgements}
\label{sec:acknowledgements}

The first author gratefully acknowledges the support of the EPSRC funded EP/L015692/1 STOR-i Centre for Doctoral Training. This work was supported by EPSRC grant EP/K014463/1, ONR Grant N00014-15-1-2380 and NSF CAREER Award IIS-1350133.

\bibliography{submission}

\begin{thebibliography}{}

\bibitem[\protect\astroncite{Abraham et~al.}{1988}]{Abraham1988}
Abraham, R., Marsden, J.~E., and Ratiu, R. (1988).
\newblock {\em Manifolds, Tensor Analysis, and Applications}, volume~2.
\newblock Springer-Verlag.

\bibitem[\protect\astroncite{Ahn et~al.}{2012}]{Ahn2012}
Ahn, S., Korattikara, A., and Welling, M. (2012).
\newblock Bayesian posterior sampling via stochastic gradient {F}isher scoring.
\newblock In {\em Proceedings of the 29th International Conference on Machine
  Learning}, pages 1591--1598. PMLR.

\bibitem[\protect\astroncite{Allaire et~al.}{2016}]{Tensor4R}
Allaire, J., Eddelbuettel, D., Golding, N., and Tang, Y. (2016).
\newblock {\em \pkg{TensorFlow}: R Interface to TensorFlow}.

\bibitem[\protect\astroncite{Baker et~al.}{2017}]{Baker2017}
Baker, J., Fearnhead, P., Fox, E.~B., and Nemeth, C. (2017).
\newblock Control variates for stochastic gradient {MCMC}.

\bibitem[\protect\astroncite{Blackard and Dean}{1999}]{Blackard1999}
Blackard, J.~A. and Dean, D.~J. (1999).
\newblock Comparative accuracies of artificial neural networks and discriminant
  analysis in predicting forest cover types from cartographic variables.
\newblock {\em Computers and Electronics in Agriculture}, 24(3):131--151.

\bibitem[\protect\astroncite{Carpenter et~al.}{2017}]{Carpenter2016}
Carpenter, B., Gelman, A., Hoffman, M., Lee, D., Goodrich, B., Betancourt, M.,
  Brubaker, M., Guo, J., Li, P., and Riddell, A. (2017).
\newblock \pkg{Stan}: {A} probabilistic programming language.
\newblock {\em Journal of Statistical Software}, 76(1).

\bibitem[\protect\astroncite{Chang and Lin}{2011}]{Chang2011}
Chang, C.-C. and Lin, C.-J. (2011).
\newblock \pkg{LIBSVM}: A library for support vector machines.
\newblock {\em ACM Transactions on Intelligent Systems and Technology},
  2(3):1--27.

\bibitem[\protect\astroncite{Chen et~al.}{2014}]{Chen2014}
Chen, T., Fox, E.~B., and Guestrin, C. (2014).
\newblock Stochastic gradient {H}amiltonian {M}onte {C}arlo.
\newblock In {\em Proceedings of the 31st International Conference on Machine
  Learning}, pages 1683--1691. PMLR.

\bibitem[\protect\astroncite{Ding et~al.}{2014}]{Ding2014}
Ding, N., Fang, Y., Babbush, R., Chen, C., Skeel, R.~D., and Neven, H. (2014).
\newblock Bayesian sampling using stochastic gradient thermostats.
\newblock In {\em Advances in Neural Information Processing Systems 27}, pages
  3203--3211. Curran Associates, Inc.

\bibitem[\protect\astroncite{Dubey et~al.}{2016}]{Dubey2016}
Dubey, K.~A., Reddi, S.~J., Williamson, S.~A., Poczos, B., Smola, A.~J., and
  Xing, E.~P. (2016).
\newblock Variance reduction in stochastic gradient {L}angevin dynamics.
\newblock In {\em Advances in Neural Information Processing Systems 29}, pages
  1154--1162. Curran Associates, Inc.

\bibitem[\protect\astroncite{Friedman et~al.}{2001}]{Friedman2001}
Friedman, J., Hastie, T., and Tibshirani, R. (2001).
\newblock {\em The Elements of Statistical Learning}, volume~1.
\newblock Springer-Verlag.

\bibitem[\protect\astroncite{Griewank and Walther}{2008}]{Griewank2008}
Griewank, A. and Walther, A. (2008).
\newblock {\em Evaluating Derivatives: Principles and Techniques of Algorithmic
  Differentiation}, volume~2.
\newblock SIAM.

\bibitem[\protect\astroncite{LeCun and Cortes}{2010}]{Lecun2010}
LeCun, Y. and Cortes, C. (2010).
\newblock {\em {MNIST} Handwritten Digit Database}.

\bibitem[\protect\astroncite{Li et~al.}{2016}]{Li2016}
Li, W., Ahn, S., and Welling, M. (2016).
\newblock Scalable {MCMC} for mixed membership stochastic blockmodels.
\newblock In {\em Proceedings of the 19th International Conference on
  Artificial Intelligence and Statistics}, pages 723--731. PMLR.

\bibitem[\protect\astroncite{Lunn et~al.}{2000}]{Lunn2000}
Lunn, D.~J., Thomas, A., Best, N., and Spiegelhalter, D. (2000).
\newblock Winbugs - a bayesian modelling framework: Concepts, structure, and
  extensibility.
\newblock {\em Statistics and Computing}, 10(4):325--337.

\bibitem[\protect\astroncite{Nagapetyan et~al.}{2017}]{Nagapetyan2017}
Nagapetyan, T., Duncan, A., Hasenclever, L., Vollmer, S.~J., Szpruch, L., and
  Zygalakis, K. (2017).
\newblock The true cost of stochastic gradient {L}angevin dynamics.

\bibitem[\protect\astroncite{Neal}{2010}]{Neal2010}
Neal, R.~M. (2010).
\newblock {MCMC} using {H}amiltonian dynamics.
\newblock In {\em Handbook of Markov Chain Monte Carlo}, volume~1. Chapman \&
  Hall.

\bibitem[\protect\astroncite{Patterson and Teh}{2013}]{Patterson2013}
Patterson, S. and Teh, Y.~W. (2013).
\newblock Stochastic gradient {R}iemannian {L}angevin dynamics on the
  probability simplex.
\newblock In {\em Advances in Neural Information Processing Systems 26}, pages
  3102--3110. Curran Associates, Inc.

\bibitem[\protect\astroncite{\pkg{TensorFlow}
  Development~Team}{2015}]{Tensorflow2015}
\pkg{TensorFlow} Development~Team (2015).
\newblock {\em {\pkg{TensorFlow}}: Large-Scale Machine Learning on
  Heterogeneous Systems}.

\bibitem[\protect\astroncite{Plummer}{2003}]{Plummer2003}
Plummer, M. (2003).
\newblock {\em JAGS: A program for analysis of Bayesian graphical models using
  Gibbs sampling}.

\bibitem[\protect\astroncite{{\proglang{R} Development Core
  Team}}{2008}]{R2008}
{\proglang{R} Development Core Team} (2008).
\newblock {\em \proglang{R}: A Language and Environment for Statistical
  Computing}.
\newblock \proglang{R} Foundation for Statistical Computing.

\bibitem[\protect\astroncite{Ripley}{2009}]{Ripley2009}
Ripley, B.~D. (2009).
\newblock {\em Stochastic Simulation}, volume~1.
\newblock John Wiley \& Sons.

\bibitem[\protect\astroncite{Roberts and Rosenthal}{1998}]{Roberts1998}
Roberts, G.~O. and Rosenthal, J.~S. (1998).
\newblock Optimal scaling of discrete approximations to {L}angevin diffusions.
\newblock {\em Journal of the Royal Statistical Society B}, 60(1):255--268.

\bibitem[\protect\astroncite{Sato and Nakagawa}{2014}]{Sato2014}
Sato, I. and Nakagawa, H. (2014).
\newblock Approximation analysis of stochastic gradient {L}angevin dynamics by
  using {F}okker-{P}lanck equation and {I}to process.
\newblock In {\em Proceedings of the 31st International Conference on Machine
  Learning}, pages 982--990. PMLR.

\bibitem[\protect\astroncite{Teh et~al.}{2016}]{Teh2014}
Teh, Y.~W., Thi{\'e}ry, A.~H., and Vollmer, S.~J. (2016).
\newblock Consistency and fluctuations for stochastic gradient {L}angevin
  dynamics.
\newblock {\em Journal of Machine Learning Research}, 17(7):1--33.

\bibitem[\protect\astroncite{Tran et~al.}{2016}]{Tran2016}
Tran, D., Kucukelbir, A., Dieng, A.~B., Rudolph, M., Liang, D., and Blei, D.~M.
  (2016).
\newblock {\pkg{Edward}: A Library for Probabilistic Modeling, Inference, and
  Criticism}.

\bibitem[\protect\astroncite{Vollmer et~al.}{2016}]{Vollmer2015}
Vollmer, S.~J., Zygalakis, K.~C., and Teh, Y.~W. (2016).
\newblock Exploration of the (non-)asymptotic bias and variance of stochastic
  gradient {L}angevin dynamics.
\newblock {\em Journal of Machine Learning Research}, 17(159):1--48.

\bibitem[\protect\astroncite{Welling and Teh}{2011}]{Welling2011}
Welling, M. and Teh, Y.~W. (2011).
\newblock Bayesian learning via stochastic gradient {L}angevin dynamics.
\newblock In {\em Proceedings of the 28th International Conference on Machine
  Learning}, pages 681--688. PMLR.

\end{thebibliography}
\bibliographystyle{apa}

\end{document}